\documentclass[useAMS,usenatbib]{mn2e}
\usepackage{psfig}

\def\ltsima{$\; \buildrel < \over \sim \;$}
\def\lsim{\lower.5ex\hbox{\ltsima}}
\def\gtsima{$\; \buildrel > \over \sim \;$}
\def\gsim{\lower.5ex\hbox{\gtsima}}
\def\be{\begin{equation}}
\def\ee{\end{equation}}
\def\mec2{m_{\rm e}c^{2}}
\def\nub{\bar{\nu}}
\def\muc{\mu_{\rm c}}
\def\mun{\mu_{\rm n}}
\def\yp{Y_{\rm p}}
\def\ypc{Y_{\rm p,c}}
\def\ypn{Y_{\rm p,n}}
\def\mp{m_{\rm p}}
\def\me{m_{\rm e}}
\def\Pr{\rm Pm}
\def\rs{r_{\rm S}}
\def\nut{\nu_{\rm t}}
\def\vr{v_{\rm rel}}

\def\ne{n_{\rm -}}

\def\msunsec{~M_{\sun} {\rm sec^{-1}}}
\def\no{\noindent}

\newcommand{\mnras}{MNRAS }

\newcommand{\aap}{A\&A }
\newcommand{\apj}{ApJ }

\newcommand{\apss}{Ap\&SS}
\begin{document}
\title[Microphysics of hyper-accreting discs]{Microphysical dissipation, turbulence and magnetic fields in  hyper-accreting discs} 

\author[Rossi, Armitage \& Menou] {Elena M. Rossi$^{1,2}$,
Philip J. Armitage$^{1,3}$ \& Kristen Menou$^4$\\
$^1$JILA, University of Colorado at Boulder, 440 UCB, Boulder, CO 80309-0440 \\
$^2$Chandra Fellow \\
$^3$Department of Astrophysical and Planetary Sciences, University of Colorado \\
$^4$Department of Astronomy, Columbia University, 550 West 120th Street, New York, NY\\
\tt e-mail:  emr@jilau1.colorado.edu (EMR); pja@jilau1.colorado.edu (PJA); kristen@astro.columbia.edu (KM)}

\maketitle
  
\begin{abstract}
Hyper-accreting discs occur in compact-object mergers and in collapsed cores of 
massive stars. They power the central engine of $\gamma$-ray bursts in most scenarios.
We calculate the microphysical dissipation (the viscosity and
resistivity) of plasma in
these discs, and discuss the implications for their global
structure and evolution.
At the temperatures ($k_{\rm b}T > \mec2$)
and densities ($\rho\sim 10^{9}-10^{12}~ {\rm gr~cm^{-3}}$)
characteristic of the neutrino-cooled innermost regions, the viscosity is provided mainly by
mildly degenerate electrons, while the resistivity
is modified from the Spitzer value due to the effects of both relativity and 
degeneracy. Under these conditions the magnetic Reynolds number is 
very large ($R_{\rm eM} \sim 10^{19}$) and the plasma behaves as an 
almost ideal magneto-hydrodynamic (MHD) fluid. Among the possible non-ideal 
MHD effects the Hall term is relatively the most important, while 
the magnetic Prandtl number, $\Pr$ (the
ratio of viscosity to resistivity), is typically larger than unity:
$10 \lsim \Pr  \lsim 6 \times 10^3$. Inspection of the outer radiatively inefficient regions indicates 
similar properties, with magnetic Prandtl numbers as high as $\sim 10^{4}$. Numerical 
simulations of the magneto-rotational instability (MRI) indicate that the saturation
level and angular momentum transport efficiency may be
greatly enhanced at high Prandtl numbers. If this behaviour persists in the 
presence of a strong Hall effect we would expect that hyper-accreting
discs should be strongly magnetised and highly variable. The expulsion
of magnetic field that cannot be dissipated at small scales may
also favour a magnetic outflow. We note that there are limited 
similaries between hyper-accreting discs and X-ray binary discs -- 
which also have a high magnetic Prandtl number close to the black hole -- which 
suggests that a comparison
between late-time activity in $\gamma$-ray bursts and X-ray binary 
accretion states may be fruitful. More generally, our results imply that the
possibly different character of high Prandtl number MHD flows
needs to be considered in studies and numerical simulations
of hyper-accreting discs.
\end{abstract}


\section{Introduction}
The fluid dynamics of dense accretion discs is described by the
Navier-Stokes and magneto-hydrodynamic equations.  Within such a
description, the microscopic properties of the fluid are parameterized
(in part) via the kinematic viscosity $\nu$, which describes the rate
at which inter-particle collisions damp fluid motions, and the
resistivity $\eta$, which quantifies the dissipation of currents via
Ohmic losses. Timescales can be associated with each of these
quantities. The viscous timescale of a disc $t_{\rm acc} = r^2 / \nu$, for
example, is the timescale on which the microscopic viscosity would
lead to angular momentum transport over a scale $r$. As is well known
\citep{pringle81}, for astrophysical discs the timescales for
evolution driven by the microscopic viscosity are many orders of
magnitude larger than the observed evolution timescales. It is
therefore generally assumed that evolution is driven by an
``anomalous" or ``turbulent" viscosity $\nu_{\rm t} \gg \nu$
\citep{shakura73}, possibly augmented in some cases by angular
momentum loss in a wind \citep{blandford82}. In most astrophysical
discs $\nu_t$ is likely generated from turbulence driven by the
magneto-rotational instability (MRI; see \citet{balbus98} and
references therein). Since $\nu_{\rm t} \gg \nu$ most studies of discs
make the implicit assumption that the actual values of $\nu$ and
$\eta$ are unimportant, and consider instead evolution under the
action of the effective transport coefficients $\nu_t$ and $\eta_t$.

Provided that the Reynolds ($R_{\rm e}$) and magnetic Reynolds ($R_{\rm eM}$)
numbers are large enough, the absolute value of either one may indeed
be of little import for accretion discs.  Their ratio, however, which
we define as the magnetic Prandtl number,
\begin{equation}
 {\rm Pm} \equiv \frac{\nu}{\eta}
\end{equation}
is almost certainly {\em not} ignorable \citep{balbus98}. An accretion
disc serves to convert the large-scale energy present in the shear
flow into other energy forms -- turbulent kinetic energy, magnetic
field, and heat -- and the dissipation required to generate heat may
vary strongly depending upon the microphysical transport
coefficients. In particular, if ${\rm Pm} \gg 1$ the plasma is highly
viscous on the scale at which resistivity operates. In this
regime one may expect that dissipation of the magnetic field will be
suppressed, with a resulting modification in the saturation level of
dynamo generated fields and an inverse cascade of magnetic energy to
large scales \citep{brandenburg01}. The importance of the magnetic
Prandtl number for the resulting magnetic field structure has been
illustarted analytically \citep{umurhan07,umurhan07b} and
numerically, both in idealised simulations of dynamo action within
forced turbulence \citep{schekochihin04} and in local shearing-box
simulations of the MRI within accretion discs
\citep{lesur07,fromang07}\footnote{Since dissipation in discs
typically occurs on small scales for which the fluid is unaware of the
large scale shear, one would anticipate that the basic physics
of large ${\rm Pm}$ dynamos is similar. Disc simulations are
required, however, to assess the impact of the small-scale physics on
MRI saturation and the large scale flow properties. It is unclear
whether the shearing-box approach is adequate for study of these
effects.}.  For discs the current limited set of simulations
indicates that the efficiency of angular momentum transport,
parameterized via the Shakura-Sunyaev $\alpha$ parameter, increases
with ${\rm Pm}$, and can approach $\alpha \sim 1$ given the
combination of modest ${\rm Pm} = 8$ and a weak net vertical magnetic
field. Motivated by these results, \citet{balbus08} calculated the
expected radial variation of ${\rm Pm}$ within standard
Shakura-Sunyaev models of geometrically thin, radiatively efficient
accretion discs. They found that ${\rm Pm}$ exceeds unity within $\sim
50$ Schwarzschild radii of the central object, and suggested that the
qualitatively different behaviour of the MRI in the high ${\rm Pm}$
limit might be responsible for the time-dependent outbursts and state
changes that are observed in X-ray binaries.

In this paper, we compute the magnitude and radial dependence of the
magnetic Prandtl number in  hyper-accreting discs.
 Such
discs, which can support accretion rates of the order of $1 \ M_\odot
\ {\rm s}^{-1}$, form from compact object mergers
\citep[e.g.][]{ruffert99} and when the cores of rapidly rotating
massive stars collapse \citep{woosley93}. 
The disc structure is shaped by neutrino emission in the regions close to the accreting object,
while in the outer parts neither neutrinos (the temperature is too low) nor photons (which are trapped within the flow)
can provide cooling.
Hyper-accreting discs are
invoked to provide the initial energy injection in many models for
$\gamma$-ray bursts (GRBs). Our ultimate motivation for studying the
magnetic Prandtl number in these discs is to understand the complex
phenomenology (jets, highly variable prompt emission,  X-ray flares, plateau phases etc) that is observed in
GRBs \citep[e.g.][]{burrows05} and which may well derive from the
evolution of the disc.  

We anticipate different results  from those
found by \citet{balbus08} for photon-cooled discs.
In the neutrino-cooled region the differences arise from two
reasons. First, since neutrino opacities are vastly smaller than
photon opacities, hyper-accreting discs are much cooler and denser
than an extrapolation of Shakura-Sunyaev discs would imply, and this
on its own would alter the magnetic Prandtl number.  Second, the
temperatures and densities in hyper-accreting discs are such that the
electrons are mildly relativistic and mildly degenerate, while the
elastic collision cross-section for the nuclei exceeds the Coulomb
cross-section. As a consequence, the classical \citet{spitzer62}
expressions for $\nu$ and $\eta$, which are adequate for ordinary
accretion flows, no longer apply. 
They do apply, instead, in the outer
regions of hyper-accreting discs, where a different $\Pr$ behaviour occurs
since the flow is radiatively inefficient and photon-pressure dominated.

The organization of this paper is as follows. In \S2 through \S5 we 
describe the formalism for calculating the structure of hyper-accreting 
discs. This is required here in order to determine consistently the plasma conditions 
within these discs -- including the temperature, density, nuclear composition 
and degree of degeneracy. Readers familiar with vertically averaged hyper-accreting 
disc models will find that our approach generally follows standard practice. 
The structure of the resulting disc models is 
presented in \S6. Given these conditions, we then calculate in \S7 and \S8 
the resisitivity and viscosity in the neutrino cooled inner 
region of the disc. These Sections contain the 
principle new results of this paper. Combining the viscosity and 
resistivity, we show in \S10 that ${\rm Pm} > 1
~{\rm or} \gg 1$ across the entire radial range relevant for GRB
models. In \S11 we estimate the strength of other non-ideal MHD 
effects, and in \S12 we study the structure of the outer non-radiative 
region of the disc. \S13 and \S14 discuss and summarize our results, 
and what they may imply for disc evolution and GRB observables.

%------------------------------------------------------------------------------------------------------------
\section{The disc model}

We calculate the temperature and density at the midplane of a
hyper-accreting disc from vertically averaged thin disc solutions.  In
hyper-accreting discs the diffusion time for photons to leak out from
the disc is much larger than the accretion time scale ($t_{\rm
diff}/t_{\rm acc} \sim 10^{11}$) and photons remain trapped. However,
for sufficiently high accretion rates\footnote{\citet{cb07} estimate a
threshold value of $\dot{M} \gsim 0.07$ M$_{\sun}$ sec$^{-1}$ for a
non-rotating black hole and $\alpha =0.1$.} the temperature in the 
innermost region exceeds $\sim$ 1~MeV and neutrino production switches
on (Popham, Woosley \& Fryer 1999).  This happens for radii r $\lsim
100~\rs$, where $\rs = 2\,GM_{\rm BH}/c^2$ is the Schwarzschild
radius for a black hole of mass $M_{\rm BH}$.  The neutrino flux
liberates the energy deposited locally by viscous (turbulent) stresses
and consequently the local sound speed $c_{\rm s}$ remains smaller
than the rotational velocity ($v_{\phi}$) of the disc. Such discs are
geometrically thin (the vertical scale height is small compared to the
radius), and can be described by a modified version of the standard
``$\alpha-$disc" theory (Shakura \& Sunyaev 1973). Many previous
studies of hyper-accreting discs have adopted this formalism
\citep[e.g.][and references
therein]{popham99,narayan01,kohri02,dimatteo02,cb07}.  In what follows
we consider only the innermost region ($r \leq 70~ \rs$) where neutrino
cooling is generally efficient. For simplicity the accretion rate is
assumed to be time-independent, but our basic results for the magnetic
Prandtl number would carry over locally to real discs in which the
accretion rate varies with radius and time.

In a steady state, conservation of mass and angular momentum for a 
differentially rotating fluid yield a relation between the 
surface density $\Sigma$ and the radial mass flow $\dot{M}$,
\begin{equation}
\Sigma\,\nut = \frac{\dot{M}}{3\,\pi}\,\left[1-\left(\frac{r_*}{r}\right)^{1/2}\right],
\label{eq:ang_con}
\end{equation} 
where the fluid equations have been averaged in the vertical
($\vec{z}$) direction over the pressure scale height $H = - P/(dP/dz)$.
We assume that no viscous torque is acting at the location of the
innermost stable orbit $r_* = 3~\rs$. Consistent with the thin
disc approximation radial pressure gradients have been neglected and
the orbital velocity is simply $v_{\phi}^2 = GM_{BH} / r$ (the ``Keplerian
velocity"). The surface density is related to the midplane ($z=0$)
density via $\Sigma = \rho H$.  The radial mass flow $\dot{M}$ [gr
s$^{-1}$] is defined by the continuity equation \be \dot{M} = - 2 \pi r
\Sigma v_{\rm r} = {\rm constant}, \ee where the radial drift velocity
$v_{\rm r}$ is subsonic.  The turbulent kinematic viscosity is
parameterized as \citep{shakura73,lynden74} , \be \nut = \alpha\,c_{\rm
s}\,H.  \ee We emphasise here that the kinematic viscosity $\nut$ is
quite different in nature from the ``microscopic" (also referred to as
``molecular") viscosity $\nu$, that enters into the definition of the
magnetic Prandtl number. The first is associated with turbulent flow
and magnetic stresses in the plasma, while the second is given by
deflection of ions and electrons by particle-particle collisions in the plasma.

For a thin disk both radial advection of energy and flux in the 
radial direction can typically be neglected (we justify this 
for our case later). From energy conservation the heating dissipated
locally by viscous stresses is emitted per unit area through the
surface of the disc as neutrino $F_{\nu}$ and anti-neutrino flux
$F_{\bar{\nu}}$ :
\begin{equation}
\frac{9}{8}\,\nu\,\Sigma\,\Omega^{2} = F_{\nu}+F_{\bar{\nu}}.
\label{eq:ene_con}
\end{equation}
The calculation of the flux in the optically thin and optically thick
regions of the disc is detailed in the next section.
Eq.~\ref{eq:ang_con} and eq.~\ref{eq:ene_con} allow us to solve for
the disc structure once the equation of state is specified.  In the
inner regions that we are considering, densities ($\rho \sim
10^{8}-10^{11}$ gr cm$^{-3}$) are high for the temperatures of
interest ($T \simeq$ a few $10^{10}$ K). This has a number of
consequences.  First, electrons are mildly degenerate ${\mu} /
{\theta} \sim {\rm a~few}$, where $\mu$ is the chemical potential in
units of the electron rest mass energy $m_{\rm e}c^2$ and $\theta =
{k_{\rm b}T} / ({m_{\rm e}c^2})$, where $k_{\rm b}$ is the Boltzmann
constant. Second, the plasma is neutron rich. If $Y_{\rm p}$ is the
number fraction of protons over baryons, $Y_{\rm p} \sim
0.1$. Finally, the baryon pressure
\begin{equation}
P_{\rm b}= \frac{\rho}{m_{\rm p}}\,k_{\rm b} T,
\label{eq:state_b}
\end{equation}
dominates the total pressure $P$ \citep[e.g.][]{belo03}, and the
isothermal sound speed $c_s^2 = P / \rho$ is, \be c_{\rm s}
=1.6 \times 10^{9} \left(\frac{T}{3 \times 10^{10}} \right)^{1/2} {\rm
cm/s}.  \ee At larger radii, on the other hand, the density is
smaller, ${\mu} / {\theta} \lsim 1$, and electron and photon
pressure are more important \citep[e.g.][]{narayan01}.  Note that in
eq.~\ref{eq:state_b}, and henceforth, we ignore the mass difference
between protons and neutrons. We also assume that all the baryons are
free, since $\alpha$ particles present at larger radii have been
photo-dissociated.  This is a valid approximation since, for the
parameters we use in our disc structure models, $\alpha$ particles do
not constitute more than 10\% of the baryons \citep[see
e.g.][thereafter CB07]{cb07}.

The neglect of advection in eq.~\ref{eq:ene_con} is generally a 
good approximation. 
The ratio of neutrino diffusion time to accretion time is
\begin{eqnarray}
\frac{t_{\rm diff,\nu}}{t_{\rm acc}} &=& \frac{\tau_{\nu}H}{c} \,
\frac{\nut}{r^{2}} \nonumber \\
& \simeq & 6 \times 10^{-4} \tau_{\nu} \left(\frac{\alpha}{0.1}\right)
\left(\frac{H/r}{0.3}\right)^{3}
\left( \frac{r / \rs}{10} \right)^{-1/2}
\end{eqnarray}
and hence advection is important only if the optical depth to
neutrinos is of the order of $10^3$ or higher. In our models optical
depths are, at most, $\tau_{\nu}\sim$ a few tens, unless the accretion
rate is very large, (e.g. $\dot{M} \gsim 10~ M_{\sun}$sec$^{-1}$, for
$\alpha=0.1$).  For a maximally rotating Kerr black hole (not treated
here), advection can become important for lower accretion rates:
$\dot{M}\gsim 2~ M_{\sun}$sec$^{-1}$, for $\alpha=0.1$ (CB07).
Advection of energy in {\it photons} instead can play a role in the
region we consider for $\alpha \gsim 0.1$, since the outer radiatively
inefficient part of the disc can extend inward, within $70~\rs$.  We will comment on this in \S~\ref{results_Trho}.

%----------------------------------------------------------------------------------------
\section{The neutrino flux}
At the temperatures and densities of interest for the inner regions 
of hyper-accreting discs neutrinos are produced primarily by electron capture onto protons
\begin{equation}
 p+e^{-}\rightarrow n+\nu
 \label{eq:pe}
 \end{equation}
while anti-neutrinos are
produced by positron capture onto neutrons 
\begin{equation}
n+e^{+}\rightarrow p+\bar{\nu}.
 \label{eq:ne}
 \end{equation} 
The inverse of these processes -- namely neutrino absorption by nuclei -- constitute 
one source of neutrino opacity, others (discussed in \S~\ref{sec:opacity}) are 
neutrino scattering by both nuclei and electrons. We note that  
at threshold reaction \ref{eq:pe} can proceed if the electron has an energy 
equal to the mass difference between a neutron and a proton. In this 
case the emitted neutrino has zero energy\footnote{A positron, instead,
can be captured by a neutron at zero energy, yielding a neutrino 
with non-zero energy. In the non degenerate limit, this
asymmetry favours an equilibrium shifted towards proton richness (see
\S~\ref{sec:nuc_comp}).}. The mild degeneracy of electrons suppresses the pair annihilation
channel $e^{+}+e^{-} \rightarrow \nu+\bar{\nu}$, since positrons are
exponentially reduced with respect to electrons:
\be
\frac{n_{+}}{n_{-}} \propto \exp{\left[-\frac{2\mu}{\theta}\right]} \ll 1.
\label{eq:suppress}
\ee
At equilibrium and neglecting advection, the capture rates
 ~\ref{eq:pe} and~\ref{eq:ne} should be equal (see also
 \S~\ref{sec:nuc_comp}). We therefore set $F_{\nu}=F_{\nub}$ and derive only the neutrino flux.  
 The calculation basically follows the CB07 prescription.

For $e^{-}$ capture onto protons (\ref{eq:pe}), the emissivity in units of $m_{\rm e}c^2$ is,
\begin{equation}
\epsilon_{\nu}=k\rho_{\rm p}\int_{Q}^{\infty} (e-Q)^3 F(e,\mu_{-})e^2 \sqrt{1 - \frac{1}{e^2}}\, de,
\label{eq:ep_n}
\end{equation}
(e.g. Shapiro \& Teukolsky 1983), where $e$ is the electron energy in units of $m_{\rm e}c^2$,  
$Q ={(m_n - m_p)} / {m_e}$, the matter
density of protons is defined as $\rho_{\rm p} = \yp \rho$, and 
$k =6.5 \times 10^{-4}$ is a constant. The function $F(e,\mu)$ is the Fermi-Dirac
distribution,
\begin{equation}
F(e,\mu) = \frac{1}{\exp[{\left(\frac{e-\mu}{\theta}\right)}] + 1}.
\label{eq:fermi}
\end{equation}
The energy density in units of $m_{\rm e}c^2$ for an
optically thin disc can be expressed in terms of the emissivity as
\begin{equation}
 U_{\nu,n} = \epsilon_{\nu}\frac{H}{c}.
\label{eq:un}
\end{equation}
If, conversely, the optical depth $\tau_{\nu}\gg 1$, neutrinos thermalize to a
Fermi-Dirac distribution (eq.~\ref{eq:fermi}) and the energy density in units of $m_{\rm e}c^2$ is given by,
\begin{equation}
 U_{\nu,c} = \frac{(m_{\rm e}\,c)^3}{2\pi^2 \hbar^3}\int_{o}^{\infty} F(e_{\nu}, \mu_{\nu})\, e_{\nu}^3 \,de_{\nu}.
\end{equation}
For $\mu_{\nu}=0$, $U_{\nu,c} = 24.5\,k_{\rm b}T^{4}$. To deal with regions of the disc 
that have $\tau_{\nu} \sim 1$ we interpolate between the optically thin and thick 
limits. We define 
\be
 x = \frac{U_{\nu,n}}{U_{\nu,n}+U_{\nu,c}}
 \label{eq:x}
\ee
and express the neutrino flux for arbitrary optical thickness as
 \be
  \frac{F_\nu}{\mec2} =\left\{ \begin{array}{ll} 
          U_{\nu,n}\, c(1+\tau_\nu)^{-1} &  \mbox{for $x<\frac{1}{2}$}, \\ 
          U_{\nu,c} \,c\,(1+\tau_\nu)^{-1} &  \mbox{for  $x\geq\frac{1}{2}$}.
                \end{array}  
        \right. 
\label{eq:flux}        
\ee

%----------------------------------------------------------------------------------------
\section{Nuclear composition}
\label{sec:nuc_comp}
In the mid-plane of the disc pairs, baryons (free protons and neutrons only in our 
case) and photons are in thermodynamic equilibrium. For given ($\rho$, $T$), and 
specified optical depth, this 
fact suffices to determine the nuclear composition of the plasma. In practice we compute the 
composition in the optically thick and optically thin limits separately, and deal with 
the intermediate regime by making an interpolation that is consistent with charge conservation. 

Let us first note some general considerations. In addition to the
processes of electron / positron capture onto nuclei
(equations~\ref{eq:pe} and~\ref{eq:ne}), pairs are created and
absorbed via \be e^{+}+e^- \leftrightarrow \gamma+\gamma.  \ee This
implies that \be \mu=\mu_{-} = -\mu_{+}.  \ee Baryons, instead, are
created and destroyed only according to the processes given by
equations~\ref{eq:pe} and~\ref{eq:ne}.  Therefore, at equilibrium
(${d\yp} / {dt}=0$), and neglecting advection, the emissivity for
neutrinos and anti-neutrinos must everywhere be equal.

In regions of the disc that are very optically {\it thick} to neutrinos,
they will have relaxed into a Fermi-Dirac distribution and
the above condition can be expressed as,
\be
 n_{\nu}=n_{\nub},
 \label{eq:eqthick}
\ee
where the number densities are calculated by integrating eq.~\ref{eq:fermi} 
over momenta with the statistical weight set to unity. 
The equality of the neutrino and anti-neutrino number densities (eq.~\ref{eq:eqthick}) implies 
a vanishing neutrino chemical potential,
\be
 \mu_{\nu} = \mu_{\nub} = 0.
 \label{eq:munu}
\ee
The chemical balance of reactions~\ref{eq:pe} and~\ref{eq:ne} (and their inverses)
 $\mu_{\rm p}+\mu = \mu_{n}+\mu_{\nu}$ gives
\be
 \mu= \theta \ln{ \frac{n_{\rm n}} {n_{\rm p}}} +Q,
 \label{eq:muthick}
\ee
\citep{belo03} where protons and neutrons (with number density $n_{\rm p}$, $n_{\rm n}$ respectively) 
have a Maxwellian distribution.

 For regions of the disk that are optically {\it thin}, we use 
\begin{equation}
 \dot{n}_{\nu} = \dot{n}_{\nub},
 \label{eq:ndotnu_ndotbnu}
\end{equation}  
 
 \no
 where the number of neutrinos $\nu$ and anti-neutrinos $\nub$ emitted per unit volume per unit time is  

\begin{equation}
\dot{n}_{\nu}=k\rho_{\rm p}\int_{Q}^{\infty} (e-Q)^2 F(e,\mu_{-})\,e^2 \sqrt{1 - \frac{1}{e^2}}\, de,
\end{equation}

\no
and 
\begin{equation}
\dot{n}_{\bar{\nu}}=k\rho_{\rm n}\int_{1}^{\infty} (e+Q)^2 F(e,\mu_{+})\,e^2 \sqrt{1 - \frac{1}{e^2}}\, de,
\end{equation}

\no
(e.g. Shapiro \& Teukolsky 1983).

In both regimes, we impose the conditions of charge neutrality,
\begin{equation}
 n_- + n_+ = Y_p \frac{\rho}{m_p},
 \label{eq:cc}
\end{equation}
where 
\begin{equation}
Y_p = \frac{n_p}{n_b},
\end{equation} 
 and baryon conservation,
\begin{equation}
 1 - \frac{n_n}{n_p+n_n} = Y_{\rm p}.
 \label{eq:bc}
\end{equation}  
The electron $n_-$ and positron $n_+$ number densities are obtained by directly integrating 
eq.~\ref{eq:fermi} over momenta with the statistical weight set to two.
To summarise, we use two different sets of equations in the two extreme regimes:
when $\tau_{\nu} \gg 1$, we find $\yp=\ypc$ and $\mu=\muc$ from 
eqs.~\ref{eq:muthick},~\ref{eq:cc} and~\ref{eq:bc}; when $\tau_{\nu} \ll 1$,
we find $\yp=\ypn$ and $\mu=\mun$ from eqs.~\ref{eq:ndotnu_ndotbnu},~\ref{eq:cc} and~\ref{eq:bc}.

Matter in the disc plane, of course, can have an arbitrary optically
thickness. To compute the nuclear composition for specified $\rho$ and $T$ 
we first interpolate to find the chemical potential using, 
\be
\mu = (1-x) \mun+x \muc,
\label{eq:mutot}
\ee
where $x$ is given by eq.~\ref{eq:x}. We then find the proton fraction 
$\yp$ by imposing charge conservation (eq.~\ref{eq:cc}). 
The resulting $\yp$ is $\ypn \le \yp \le \ypc$.

Physically, the most important trends in the nuclear composition can be 
understood in terms of the ratio of the temperature to the (relativistic) 
degeneracy temperature 
\be
T_{\rm deg} \simeq  0.71 \left(\frac{\rho \yp}{m_{\rm p}}\right)^{1/3}  \rm K,
\label{eq:tdeg}
\ee
where we set $n_{\rm p} = n_{\rm -} = \rho \yp/m_{\rm p}$. If 
$T < T_{\rm deg}$ (at low temperature, or at high density) then $\mu$ increases, 
positrons are suppressed (eq.~\ref{eq:suppress}), and the resulting plasma 
is neutron-rich (low $\yp$). In the opposite limit of $T \gg T_{\rm deg}$, $n_{+} \simeq n_{-}$ but  $\yp \gsim 0.5$  
because, as noted before, reaction \ref{eq:ne} is energetically favoured.

%---------------------------------------------------------------------------------------------------------------------
\section{Opacities}
\label{sec:opacity}

The vertical optical thickness of the disk to $\nu$ and $\nub$ in eq.~\ref{eq:flux}  

\be
\tau_{\nu} = H \rho \kappa,
\label{eq:tau}
\ee

\no
is calculated considering three sources 
of opacity: $\kappa = \kappa_{\rm a}+\kappa_{\rm s,e}+\kappa_{\rm s,b}$ [cm$^2$ gr$^{-1}$].
First, absorption onto nuclei,

\be
\kappa_{\rm a} = \frac{0.5}{m_{\rm p}} \left(\sigma_{\nu,\rm n}(1-\yp)+\sigma_{\nub,\rm p}\yp\right) 
\ee 

\no
where $\sigma_{\nu,\rm n(p)}$ is the cross section for $\nu$ ($\nub$) absorption by neutrons (protons).
Second, elastic scattering by electrons,

\be
\kappa_{\rm s,e} = 0.5 \frac{Y_{\rm e}}{m_{\rm p}} \left(\sigma_{\rm e,\nu}+\sigma_{\rm e,\nub}\right),
\ee

\no
where $Y_{\rm e} = {n_{-}}/{n_{\rm b}}$ and  $\sigma_{\rm e,\nu(\nub)}$ is 
the electron-neutrino (anti-neutrino) cross section. 
We neglect scattering onto positrons since $n_{+}/n_{-} \ll 1$. 
Finally, elastic scattering by nuclei,

\be
\kappa_{\rm s} = \left(0.3~\yp+0.36~(1-\yp) \right) \frac{\sigma_0} {m_{\rm p}} \,\bar{E}_{\nu}^{2},
\ee

\no
where  $\sigma_0 = 1.7 \times 10^{-44}$ cm$^{2}$. 
For the explicit forms of the cross sections, we refer the reader to the Appendix B of CB07.

The cross sections for these processes depend on the neutrino
energy, and the total cross section $\sigma$ should be derived
by integrating over the neutrino spectrum. This requires knowledge of the 
neutrino distribution function at arbitrary optical depth. We, instead,
use approximate mean cross sections, which are adequate for our purposes. We
first evaluate $\sigma$ at a mean neutrino energy (in unit of $\mec2$) :
$$\sigma = \sigma(\bar{E}_{\nu}).$$
Then, the usual interpolation allows us to link the optically thin and thick regimes:

\be
\bar{E}_{\nu} = (1-x)\,\bar{E}_{\nu,\rm n}+x\,\bar{E}_{\nu,\rm c},
\label{eq:emean}
\ee

\no
(CB07) where 

\be
\bar{E}_{\nu,\rm n} = \frac{\epsilon_{\nu}}{\dot{n}_{\nu}},
\ee

\no
corresponding to $\tau_{\nu} \ll 1$ and

\be
\bar{E}_{\nu,\rm c} \simeq 3.17\;\frac{k_{\rm b}T} {\mec2} ,
\ee

\no corresponding to $\tau_{\nu} \gg 1$ is the mean energy calculated
from a Fermi-Dirac distribution function with zero chemical potential.
We use the same treatment for anti-neutrino cross sections.

In practice, we find that for conditions appropriate to our disc
models absorption by nuclei is always the dominant opacity by one or
two orders of magnitude. Electron scattering is least important.
%----------------------------------------------------------------------------------------
\section{Results for the disc structure}
 \label{results_Trho}
%%%%%%%%%%%%%%%%%%%%%%%%%%%%%%%%%%%%%%%%%%%%%%%%%%%
\begin{figure}
\psfig{figure=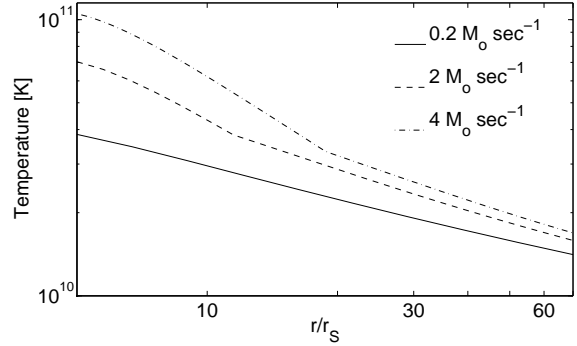,width=0.48\textwidth}
\caption[]{Midplane temperature as a function of radius (in
units of the Schwarzschild radius $\rs$), for three disc models
computed for $M_{\rm BH} =3\, M_{\sun}$ and $\alpha = 0.1$.  The
accretion rates are indicated in the legend.}
\label{fig:T_r}
\end{figure}
\begin{figure}
\psfig{file=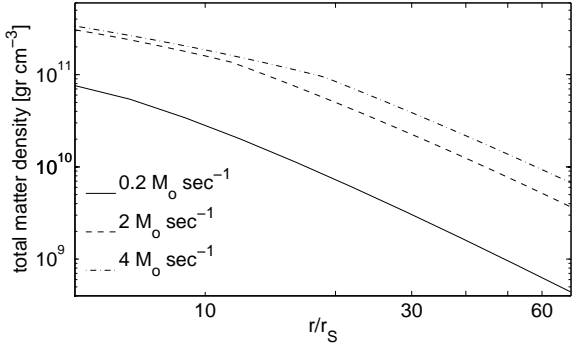,width=0.48\textwidth}
\caption[]{The same as fig.~\ref{fig:T_r} but for the midplane total matter density.}
\label{fig:rho_r}
\end{figure}

 
%%%%%%%%%%%%%%%%%%%%%%%%%%%%%%%%%%%%%%%%%%%%%%%%%%%%
To give an idea of the midplane temperatures and densities encountered
in hyper-accreting discs, we show in Figure~\ref{fig:T_r} and
Figure~\ref{fig:rho_r} illustrative examples of the radial profiles of
these quantities. The disc models plotted were computed for a black
hole of $3~$M$_{\sun}$, with $\alpha =0.1$ and accretion rates of
$0.2$, $2$ and $4~$M$_{\sun}$sec$^{-1}$.  For these parameters the
disc is optically thin across the entire radial range shown for the
lowest value of the accretion rate. For the two high accretion rate
models the disc is optically thick within about $15 \rs$ of the black
hole.

The most important feature of the neutrino cooled solutions for our
purposes is the relatively weak radial dependence of the
temperature. As noted elsewhere (e.g. CB07), the disc self-adjusts to
maintain electrons in a state of mild degeneracy (i.e. $T \approx 0.3
- 0.5 T_{\rm deg}$): higher degeneracy would cause the cooling to
decrease drastically and the temperature to rise, lifting the
degeneracy, and vice versa for lower degeneracy. This thermostatic
aspect of neutrino cooled discs has as a consequence that temperature
does not vary strongly with radius, since the degeneracy temperature
depends only weakly on density (and in addition $\yp$ decreases with
$\rho$).  The proton fraction $\yp$ is everywhere less than $0.3$,
falling to around $\yp\sim 0.1$ in the innermost regions.

We have verified that the physical conditions in our disc models
reproduce those of prior authors. Comparing our results to those of
CB07, we find excellent agreement -- especially for $\alpha<0.1$ --
between our models with a given $\alpha$ and their model with $\alpha
/ 2$ (the factor two arises primarily due to different definitions of
viscosity and different inner boundary conditions, both of which are
largely arbitrary). For $\alpha = 0.1$, their densities are lower and
the advection dominated outer region (which is not cooled efficiently
by neutrinos) extends to smaller radii.  Since we ignore the advective
term in the energy equation, our profiles have a different slope but
they never differ by more than a factor of 2 in density and by 50\% in
temperature.
%%%%%%%%%%%%%%%%%%%%%%%%%%%%%%%%%%%%%%%%%%%%%%%%%%%%
%%%%%%%%%%%%%%%%%%%%%%%%%%%%%%%%%%%%%%%%%%%%%%%%%%%%
%------------------------------------------------------------------------------------------------------------
\section{Electrical resistivity of the plasma}
\label{sec:eta}

The computation of the resistivity of a plasma in which 
electrons are both relativistic and degenerate is quite subtle, and 
ultimately requires numerical methods. We first review some 
general considerations, including the classical Spitzer (1962) 
result, which we emphasise is {\em not} applicable to hyper-accreting 
discs. These results will prove useful later when we discuss the 
relative importance of different non-ideal MHD effects.
%------------------------------------------------------------------------------------------------------------
\subsection{General considerations}

The electrical resistivity of a plasma is a measure of the electron
mobility in the presence of an applied electric
field. Acceleration due to the electric field ($e \vec{E}$, where 
$e$ is now the electric charge) is opposed by drag from 
(primarily) electron-proton collisions.
Balancing these competing forces on a electron of velocity $v$ and momentum
$p=\gamma_{\rm e}\,\me v$,

$$e\vec{E} =\frac{p}{t_{\rm ep}}, $$
where we neglect pressure gradients and magnetic fields and assume stationary protons. 
Ohm's law then gives,
\be
\eta = \frac{\gamma_{\rm e} } {4 \pi r_{\rm e} n_{-} t_{\rm ep}},
\label{eq:eta}  
\ee

\no (see e.g. Jackson 1963 p.~460, for the non-relativistic 
derivation). In this expression $n_{-}$ is strictly
equal to $n_{-}(p)$: the number density of electrons with momentum
$p$, $\gamma_{\rm e}$ is the electron Lorentz factor, $r_{\rm
e}=e^{2}/\mec2 \simeq 2.82 \times 10^{-13} {\rm cm}$ is the classical
electron radius and $t_{\rm ep}$ is the proton-electron deflection
time. The associated electrical conductivity is
$\sigma = c^2/(\eta 4\pi)$. Extending the non-relativistic definition (that can be found for example in  Longair 1992,
p.~304), $t_{\rm ep}$ can be defined so that $\left < \Delta
p_{\perp}^2 \right>\,t_{ep} = p^{2}$, where $\left < \Delta p_{\perp}^2
\right>$ is the mean square impulse received by an electron per
scattering perpendicular to its direction of motion. We obtain

\be
t_{\rm ep} = \frac{p^2\,v}{8 \pi n_{\rm p} e^{4} \log\Lambda_{\rm ep}(p)},
\label{eq:tep}
\ee

\no
 where 
$\log\Lambda_{\rm ep}(p)$ is the Coulomb logarithm, defined as the logarithm of the maximum to the minimum
impact parameter for proton-electron collisions.
 
 If we take the non-relativistic limit ($\gamma_{\rm e}
  \rightarrow 1$) of eqs.~\ref{eq:eta} ~and ~\ref{eq:tep} and insert 
  the mean thermal velocity of non-degenerate electrons
  $v=\bar{v}_{\rm e}=\sqrt{{8 k_{\rm b}T}/\pi {\me}}$, we recover the scalings\footnote{There 
  is a numerical factor of $3.5$
  between the Spitzer resistivity and our derivation.} derived by Spitzer
  (1962) for a completely ionized plasma with $n_{\rm p}=n_{-}$  

\be \eta_{\rm S} = 1.9 \times 10^{12}
  \log\Lambda_{\rm ep,S}\,T^{-3/2} \; {\rm \frac{cm^{2}}{sec}},
\label{eq:etas}
\ee

\no 
where the Coulomb logarithm for electron-proton collision is:
 \be
 \Lambda_{\rm ep,S} = 1.24\times 10^{4}
\frac{T^{3/2}}{\sqrt{n_{-}}}\sqrt{\frac{4.2\times 10^{5}} {T}}.
\label{eq:cou_log_ep_S}
\ee 

\no
In this non-degenerate regime, the (inverse) dependency of $\eta_{\rm S}$ 
on density is weak.

%------------------------------------------------------------------------------------------------------------
\subsection{Numerical computation of $\eta$}

To compute the plasma resistivity under our conditions of temperature and 
density, we use the code by \citet{pot99a} and \citet{pot99b}\footnote{http://www.ioffe.ru/astro/conduct/}, 
which is able to treat arbitrary degeneracy of fully relativistic electrons (but not 
positrons). The calculation involves an integration of 
eq.~\ref{eq:eta} over energy, weighted by the Fermi-Dirac distribution
(eq.~\ref{eq:fermi}).  The Coulomb logarithm is calculated via an
effective scattering potential that takes into account Debye charge
screening (see reference above for details).  We use as input
parameters to the code the temperature $T$ and the proton density
$\rho_{\rm p}$, which, in turn, depends (via $\yp$) on $T$ and $\rho$.
From $\rho_{\rm p}$ the code derives the electron density and thus the
electron chemical potential which uniquely specifies the electron
Fermi-Dirac distribution. 

In principle the calculation of the total 
resistivity  should include the contribution of positrons both as
charge carriers (they contribute to the current density and have their
own resistivity) and as deflectors for electrons.  However, we have
already noted that positrons are exponentially suppressed with
respect to electrons and from charge conservation it follows that they
can also be neglected in the drag force which remains dominated by protons.

 The code allow us to explore the plasma resistivity under different
 condition of electron degeneracy.  In the non-degenerate
 limit the resistivity {\it decreases} with temperature, because
 scattering becomes less frequent, while the density dependence is weak and confined to the
 Coulomb logarithm (e.g. Spitzer resistivity, eq.~\ref{eq:etas}, for $k_{\rm
 b}T\ll \mec2$).  In
 the highly-degenerate regime, instead, the resistivity {\it
 increases} with temperature, because more energy states become
 available for the electrons to be scattered into. The density
 dependence here is strong. Higher density (for fixed 
 temperature) results in higher degeneracy and suppression of
 scattering: the plasma resistivity thus decreases. The intermediate
 mildly degenerate regime, where the resistivity changes very slowly
 with temperature and density, is the one occupied by our plasma.
%------------------------------------------------------------------------------------------------------------
\subsection{Results}
\begin{figure}
\psfig{file=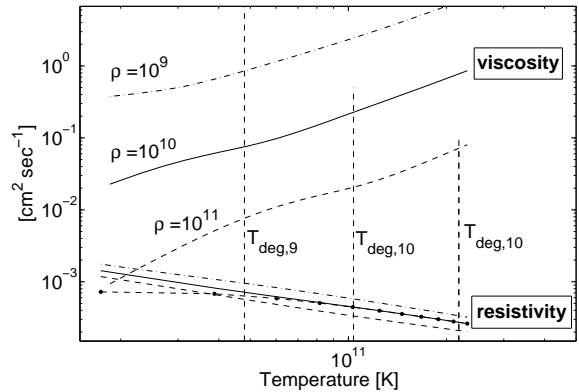,width=0.48\textwidth}
\caption[]{The microscopic viscosity (upper curves) and electrical
resistivity (lower curves) as a function of temperature, for three
different densities: $\rho = 10^{9}$ gr cm$^{-3}$ (dot-dashed line),
$\rho = 10^{10}$ gr cm$^{-3}$ (solid line) and $\rho = 10^{11}$ gr
cm$^{-3}$ (dashed line).  These curves are calculated with a
self-consistent nuclear composition and chemical potential $\mu$.  The
resistivity curve plotted with a dashed line with dots is computed for
$\rho =10^{10}$ but with a {\em fixed} nuclear composition, $\yp\simeq
0.5 $, and the corresponding $\mu$ for each temperature: this forces
the electrons to become degenerate as $T$ decreases.  The proton
fraction $\yp\simeq 0.5$ corresponds to the maximum temperature ($\sim
2\times 10^{11}$ K) considered here, for any shown density.  All
curves are computed in the optically thin limit.  The vertical lines
show the degeneracy temperature (eq.~\ref{eq:tdeg}), calculated for a
given density with $\yp\simeq 0.5$.}
\label{fig:eta_nu_T}
\end{figure}

The calculated resistivity is shown as a function of temperature 
in Figure~\ref{fig:eta_nu_T}. For a density of $\rho = 10^{10} \ 
{\rm gr} \ {\rm cm}^{-3}$ the degeneracy temperature 
$T_{\rm deg,10}$ (evaluated at $\rho_{\rm p} = 0.5 \times 
10^{10} \ {\rm gr} \ {\rm cm}^{-3}$) is approximately $10^{11}$~K. 
For temperatures above this value, the proton
fraction remains approximately constant at a value of $\yp\simeq 0.5$,
thus $\rho_{\rm p}$ does not change appreciably and the resistivity
decreases with temperature since electron-proton
collisions becomes less frequent. For lower temperatures, one's 
naive expectation is that the resistivity ought to flatten out and 
eventually decrease as electrons become more and more degenerate. 
This behaviour is seen {\em if we artificially impose} a temperature 
independent electron density (dashed line), but it does not reflect 
the physical behaviour of the disc plasma. In fact, as the temperature 
drops toward and below $T_{\rm deg,10}$ the plasma becomes more and 
more neutron-rich (i.e. $\yp$ decreases), with an attendant decrease in the 
electron density. The effective degeneracy temperature decreases, and 
the electrons never become highly degenerate. The overall result is 
that at ``low" temperatures the electrons self-adjust toward a 
mildly degenerate regime in which $\eta$ rises slowly with decreasing 
temperature. The dependence on density is weak.

\begin{figure}
\psfig{file=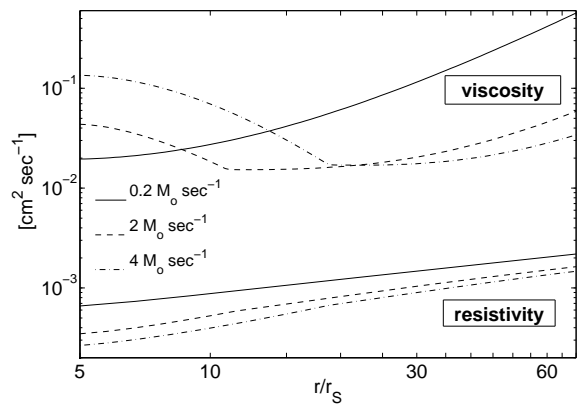,width=0.48\textwidth}
\caption[]{Microscopic viscosity (upper curves) and electrical resistivity (lower curves) as a function of
 scaled radius, for the three disc models shown in
 Figure~\ref{fig:T_r}.}
\label{fig:eta_nu_r}
\end{figure}

Within a hyper-accreting disc, the midplane temperature is a fairly 
weak function of radius (Figure~\ref{fig:T_r}). This, when coupled with 
the weak dependence of the resistivity on density, results in a 
radial profile of $\eta$, shown in Figure~\ref{fig:eta_nu_r},  
that increases slowly with distance from the black hole. Between 
5 and $70~\rs$ the variation in the resistivity is less than an 
order of magnitude. There is a modest {\em decrease} with increasing 
accretion rate.

%%%%%%%%%%%%%%%%%%%%%%%%%%%%%%%%%%%%%%%%%%%%%%%%%%%%%%%%%%%%%%%%%%%
%%%%%%%%%%%%%%%%%%%%%%%%%%%%%%%%%%%%%%%%%%%%%%%%%%%%%%%%%%%%%%%%%%%%

%%%%%%%%%%%%%%%%%%%%%%%%%%%%%%%%%%%%%%%%%%%%%%%%%%%%%%
%%%%%%%%%%%%%%%%%%%%%%%%%%%%%%%%%%%%%%%%%%%%%%%%%%%%%%%
%%%%%%%%%%%%%%%%%%%%%%%%%%%%%%%%%%%%%%%%%%%%%%%%%%%%%%%%%%%%%%%%%%%
\begin{figure}
\psfig{file=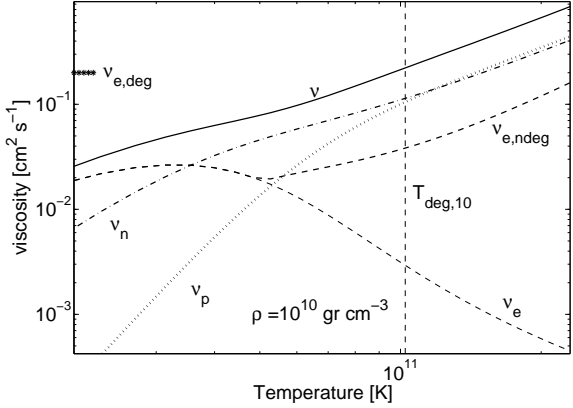,width=0.48\textwidth}
\caption[]{Viscosity components as a function of temperature for
matter density $\rho = 10^{10} {\rm gr ~cm^{-3}}$.  The corresponding
curves are labelled. The electron viscosity $\nu_{\rm e}$ appropriate 
for the partially degenerate regime is shown as the light dashed line. 
This component, which is the one that we use in the disc calculation, 
dominates the total viscosity (shown as the solid 
curve) at low temperatures. At high temperatures the viscosity 
due to non-degenerate electrons ($\nu_{\rm e,ndeg}$, shown as 
the heavy dashed line on the right) increases, but in this regime it is always 
negligible compared to the contributions from neutron-neutron 
($\nu_n$) and proton-neutron ($\nu_p$) collisions. At the upper left 
of the plot we show the value of the electron viscosity in the 
strongly degenerate regime (which we emphasize is not appropriate 
for disc conditions).}
\label{fig:nu_T_comp}
\end{figure}
%%%%%%%%%%%%%%%%%%%%%%%%%%%%%%%%%%%%%%%%%%%%%%%%%%%%%%%%%%%%%%%%%%%%
%---------------------------------------------------------------------------------------------
\section{Microscopic viscosity of the plasma}
\label{sec:nu}

Contributions to the viscosity of the plasma come from three sources, proton-proton 
Coulomb collisions, elastic baryonic scattering due to the strong force, and 
electron viscosity. It is not immediately obvious which of these processes 
dominates for a mildly relativistic and degenerate neutron-rich plasma. 
We first consider plasma viscosity due to baryon collisions.
Typically (but not in our case) only ion-ion Coulomb collisions need to be taken into account.
Protons are non-degenerate and non-relativistic, so we use the
non-relativistic limit of eq.~\ref{eq:tep}, with $p=\mp v$ where 
the mean proton relative velocity $v=\vr =\sqrt{2}\bar{v}$ is 
written in terms of the mean thermal velocity $\bar{v} = \sqrt{8 k_{\rm b} T/\pi \mp}$. 
Multiplying by $(\log \Lambda_{\rm ep,S} / \log \Lambda_{\rm pp,S})$
we obtain the proton-proton (p-p) deflection timescale,

%\be
%t_{\rm pp,S} = 16.3 \; \frac{T^{3/2}} {n_{\rm p}\log{\Lambda_{\rm pp,S}}} \;{\rm sec},
%\label{eq:tep_S}
%\ee

\be
t_{\rm pp,S} = 18.03 \; \frac{T^{3/2}} {n_{\rm p}\log{\Lambda_{\rm pp,S}}} \;{\rm sec},
\label{eq:tep_S}
\ee

\no
 where the p-p Coulomb logarithm is given by

\be
\Lambda_{\rm pp,S} = 1.24\times 10^{4} \frac{T^{3/2}}{\sqrt{n_{-}}} 
\label{eq:cou_log}
\ee 

\no
\citep{spitzer62}\footnote{Our
definition differs by a factor of $\sim 1.6$ from the
classic Spitzer result.}. Scattering in this regime is dominated 
by small-angle deflections.

 Once plasma temperatures exceed $\sim$~MeV, however, as in
hyper-accreting discs, the elastic nuclear cross section becomes
larger than the Coulomb one.  The shape of the nuclear two-body cross
section can be derived theoretically with a partial-wave analysis
(considering only s-waves) within the frame-work of the
effective-range theory \citep{bethe}.  For the particle energies of
interest to us, the {\em partial} cross section (for a {\it single}
total spin state) depends only on two parameters: the scattering
length $a$ and the effective range of the nuclear potential $r_{\rm
ef}$,

\be
\sigma_{\rm 0,i} = \frac{4\pi a_{\rm i}^{2}}{[1-\frac{1}{2}a_{\rm i}r_{\rm ef,i}\kappa^2]^2+(a_{\rm i} \kappa)^2},
\ee

\no where the wave number is $\kappa = (1/\hbar)\sqrt{2\mp E}$, where
$E = (3/4) k_{\rm b} T$ is the centre-of-mass total kinetic energy.
For the p-n collision we need to consider contributions from both
singlet $i=s$ (antiparallel spins) and triplet $i=t$ (parallel spins)
states, and the total cross section is the weighted sum of partial
siglet and triplet cross sections, \be \sigma_{\rm pn} = \frac{3}{4}
\sigma_{0,t}+ \frac{1}{4}\sigma_{0,s}, \ee with $a_{t}=5.41 \times
10^{-13}$ cm, $a_{\rm s}=-23.71 \times 10^{-13}$ cm, $r_{\rm
ef,t}=1.74 \times 10^{-13}$ cm and $r_{\rm ef,s}=2.75 \times 10^{-13}$
cm \citep{hack06}.  For identical particles, however, the triplet
state is excluded. The p-p and n-n cross section are thus, \be
\sigma_{\rm pp} = \frac{1}{4} \sigma_{0,s}, \ee with $a_{\rm s}
=-17.31 \times 10^{-13}$ cm and $r_{\rm ef,s} = 2.85 \times 10^{-13}$
cm; $\sigma_{\rm nn} = (1/4)\sigma_{0,s}$, with $a_{\rm s} =-18.8
\times 10^{-13}$ cm and $r_{\rm ef,s} = 2.75 \times 10^{-13}$ cm
\citep{miller90}.  The result is that $\sigma_{\rm pn}>\sigma_{\rm
nn}\simeq \sigma_{\rm pp}$.

Once we allow for the possibility of strong interactions  
the total collision timescale for protons becomes 

\be
\frac{1}{t_{\rm p}} = \frac{1}{t_{\rm pp,S}}+ \frac{1}{t_{\rm
pp,nu}}+\frac{1}{t_{\rm pn}} \simeq \frac{1} {t_{\rm pn}},
\ee

\no where the nuclear p-p and p-n timescales are $t_{\rm
pp,nu}=1/(\sigma_{\rm pp} n_{\rm p} \vr)$ and $t_{\rm pn}=
1/(\sigma_{\rm pn} n_{\rm n}\,\vr)$ respectively.  The collision
timescale for a neutron is instead \be \frac{1}{t_{\rm n}} =
\frac{1}{t_{\rm np}}+ \frac{1}{t_{\rm nn}}\simeq \frac{1}{t_{\rm nn}},
\ee

\no where $t_{\rm np} = 1/(\sigma_{\rm pn} n_{\rm p} v_{\rm s})$ and
$t_{\rm nn} = 1/(\sigma_{\rm nn} n_{\rm n}\,v_{\rm s})$.  Note that
for both protons and neutrons the collision timescale is dominated by
encounters with {\it neutrons}, since the flow is neutron rich and the
high temperature inhibits electromagnetic collisions. The
corresponding mean free paths are $\lambda_{\rm p} = \bar{v}t_{\rm p}$
and $\lambda_{\rm n} = \bar{v}t_{\rm n}$.

Beside baryons, we should consider the contribution to viscosity from
electrons, since degeneracy results in a longer mean free path
$\lambda_{\rm e}= \beta\,c\, t_{\rm ep}$ ($\beta$ is the ratio of
electron velocity to $c$) and a higher momentum density than
non-degenerate electrons.

According to the classical kinetic theory the dynamical viscosity
$\eta_{\rm \nu}\; {\rm [gr\; cm^{-1}sec^{-1}]}$ for a single species
``i" is proportional to $\rho_{\rm i} v_{\rm i} \lambda_{\rm i}$.
Therefore, we can write \be \eta_{\rm \nu}= C_{1} \left(\rho_{\rm
p}\bar{v}\lambda_{\rm p} +\rho_{\rm n}\bar{v}\lambda_{\rm n}+\ne p
\lambda_{\rm e}\right) \ee \no where $C_{1}$ is a constant of the
order of unity\footnote{We assume that the constant $C_{1}$ is the
same for the collision processes considered here.}.  The typical
electron kinetic energy is $\mu \me\,c^2$, where $\mu$ is the chemical 
potential, thus its momentum is
$p=\me\,c\sqrt{\mu^2+2\mu}$.

The total kinematic viscosity $\nu=\eta_{\rm \nu}/\rho$ [cm$^{2}$\,s$^{-1}$] can be written as 

\be
\nu = \nu_{\rm S} \left(\yp \frac{\lambda_{\rm p}}{\lambda_{\rm S}} +(1-\yp) \frac{\lambda_{\rm n}}{\lambda_{\rm S}}+\frac{\ne p}{\rho \bar{v}} 
\frac{\lambda_{\rm e}}{\lambda_{\rm S}} \right)
\label{eq:nu}
\ee

\no where $\rho \nu_{\rm S}= C_{1} \rho \bar{v} \lambda_{\rm S}$ and $\lambda_{\rm S} =  \bar{v} t_{\rm pp,S}$
 (where $t_{\rm pp,S}$ is calculated for $n_{\rm p}=n_- =n$)
 is the dynamical viscosity for a fluid of completely ionized hydrogen,

\be
\nu_{\rm S} = 2.21\times 10^{-15} \frac{T^{5/2}}{\rho\,\log\Lambda_{\rm pp,S}},
\label{eq:nuspitzer}
\ee
  
\no (Braginskii 1957, 1958; Spitzer 1962).  In eq.~\ref{eq:nu}, the
first term on the right is the proton viscosity $\nu_{\rm p}$, the
second is the neutron viscosity $\nu_{\rm p}$ and the third is the
electron viscosity $\nu_{\rm e}$. This last reduces formally to the formula for
strongly degenerate electrons $\rho \nu_{\rm e,deg}~ \propto \ne\; p_{\rm
F}\,v_{\rm F}\, t_{\rm ep}$ \citep{nand84} (with $C_{1} = 1/5$).

In fig.~\ref{fig:nu_T_comp}, we plot $\nu$ and its components as a
fucntion of temperature.  In the non-degenerate limit ($T>T_{\rm
deg}$), protons and neutrons contribute equally to the total
viscosity: they carry the same momentum density, transported in both
cases by n-p collisions. In this regime, the electron contribution
$\nu_{\rm e,ndeg}$ has been calculated as $\nu_{\rm e,ndeg} = \nu_{\rm
S} \left(Y_{\rm e}\,\bar{v}_{\rm e}^2\,t_{\rm ep}/
\bar{v}\,\lambda_{\rm ep,S}\right)$, where $Y_{\rm e}$ is the electron
fraction and $t_{\rm ep} = 0.15~{\rm sec}~ T^{3/2}/n_{\rm p}
\log\Lambda_{ep,S}$ (eq.~\ref{eq:tep} with $v=\bar{v}_{\rm e}$ and $p
= \bar{v}_{\rm e}\me$).  As the temperature decreases, the flow
becomes neutron rich and the electrons degenerate. As this happens the
proton momentum density drops faster than the neutron's one, strong
collisions with neutrons dominate the baryon collision frequency and
the electron mean free path increases (while the baryon one keeps
decreasing). As a consequence, electron viscosity becomes comparable
or greater than the baryon viscosity. Since the electron density
decreases with temperature, $\nu_{\rm e}$ eventually decreases (even
if $\lambda_{\rm e}$ keeps increasing) and it does not attain the
limiting viscosity value for strongly degenerate electrons (shown in
the figure as the horizontal dashed line). Therefore the total
viscosity $\nu$ {\it decreases} as the plasma cools even when electron
viscosity dominates.  The behaviour of $\nu$ as a function of density
is shown in Figure~\ref{fig:eta_nu_T}. It decreases as the plasma
becomes denser, since p-n collisions become more frequent and the
electron (and thus their momentum) density decreases when below the
nominal degeneracy temperature.  In these figures, the $T < T_{\rm
deg}$ regime represents the typical situation in hyper-accreting
discs.

Figure~\ref{fig:eta_nu_r} shows the predicted radial profile of the
viscosity within the hyper-accreting disc models.  In the optically
thin regions of the disc the viscosity increases with radius, since
the temperature in this region decreases much more slowly with radius
than the density. Electron viscosity is higher than baryon viscosity
but the two contributions tend to converge (and they can become
comparable for low accretion rates) in the less dense regions away
from the black hole where the electrons are less degenerate.  The
trend is the opposite in the optically thick innermost region, which
is present only at high accretion rates.  In this case the viscosity
decreases moving outward.  Here, the neutron contribution is highly
suppressed ($\sim$ one-two orders of magnitude) and electrons
completely dominate the flow viscosity.

%--------------------------------------------------------------------------------------------
\subsection{Neutrino and photon viscosities}

Photons and neutrinos can also in principle contribute to viscosity. 
In the case of neutrinos, the mean free path is
of course enormous compared to that of the baryons, so momentum
transport by neutrinos will dominate {\em if it acts as a
viscosity}. This will depend upon the neutrino optical depth. In
regions of the disc that are very optically thick neutrino viscosity
will act as a (very large) Navier-Stokes viscosity.  Indeed,
\cite{masada07} have suggested that neutrinos might be able to shut
off the MRI completely, by reducing the Reynolds number below that
which admits linear growth of MRI modes\footnote{Note that one cannot
easily appeal to neutrinos to shut off turbulence entirely, since the
neutrino flux itself arises as a consequence of heating driven by the
very same turbulence.}. In optically thin regions, on the other hand,
momentum transport by neutrinos is non-local and acts as a radiation
drag rather than a viscosity \citep{arav92}.  In this paper we are
principally concerned either with optically thin regions of the disc
(where neutrinos do not contribute viscosity), or with regions of
modest optical depth ($\tau_\nu \sim$~a few) for which the correct
treatment is unclear. The reader should, however, bear in mind that
the presence of neutrinos may have dynamical consequences in the
optically thick regions of the flow.  Since we derive high values 
of the Prandtl number while ignoring the neutrino contribution, to the
extent that neutrinos contribute to viscosity, they will only
reinforce our conclusions on the large Pm regime of hyper-accreting
discs. Our magnetic Prandtl numbers are conservative in that sense.

The discussion on the different regimes according to the optical
thickness also applies to photon viscosity.  Our discs are everywhere
in the mid-plane extremely optically thick to photons. This implies a
much smaller mean free path for photons than for neutrinos and their contribution to
viscosity is relatively suppressed by a factor equal to the ratio of
mean free paths for the two species.  Photon viscosity can also be
neglected with respect to baryon viscosity, even if they have similar
mean free paths, since the photon momentum density is small compared to the baryons' one.
% are less efficient at transporting
%momentum: per particle the photon viscosity is smaller by a factor $\bar{v}/c$, where
%$\bar{v}$ is the mean thermal velocity of baryons.
%--------------------------------------------------------------------------------------------------------
\section{Comparison with X-ray binary discs}
%%%%%%%%%%%%%%%%%%%%%%%%%%%%%%%%%%%%%%%%%%%%%%%%%%%%%%%%%%%%%%%%%%%
\begin{figure}
\psfig{file=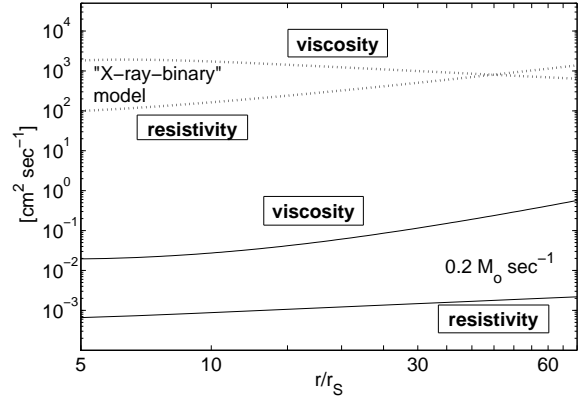,width=0.48\textwidth}
\caption[]{Comparison between the ``X-ray binary'' disc model and the
``neutrino cooled'' disc model for the electrical resistivity and the
microscopic viscosity as a function of scaled radius. The
``X-ray binary'' disc has a $10 M_{\sun}$ central object, $\dot{M} =
0.1 L_{\rm edd}/c^2$ (where the Eddington luminosity is $\sim 10^{39}$
erg s$^{-1}$) and $\alpha = 0.01$.  The neutrino cooled disc model has
$\dot{M} = 0.2\,M_{\sun}$sec$^{-1}$, $\alpha = 0.1$ and $M_{\rm BH}=
3\, M_{\odot}$.}
\label{fig:eta_nu_rb}
\end{figure}
%%%%%%%%%%%%%%%%%%%%%%%%%%%%%%%%%%%%%%%%%%%%%%%%%%%%%%%%%%%%%%%%%%%

Identical methods can be used to compute the resistivity and 
viscosity for photon-cooled discs, which are present in X-ray 
binaries at similar radii around black holes of a few Solar masses. 
In Figure~\ref{fig:eta_nu_rb} 
we plot the radial profiles of these quantities for a model 
presented by Balbus \& Henri (2007) as representative of a X-ray binary 
disc. The model has a central black hole of mass $10 \ M_\odot$, 
surrounded by a disc accreting at one tenth the Eddington limit with 
$\alpha = 0.01$. We treat the photon flux in the diffusion 
approximation, adopt an opacity that is of Kramers' form plus 
an electron scattering correction ($\kappa \simeq 0.43+6.6~ \rho T^{-7/2}$), 
and write the total pressure as the sum of the gas pressure 
(baryons and electrons) and photon pressure. The abundances 
are taken as 10\% $\alpha$ particles and 90\% hydrogen by number. 
The gas is completely ionised. In the X-ray binary regime the 
Spitzer resistivity (eq.~\ref{eq:etas}) is valid (we neglect $e-\alpha$ collisions),
while the viscosity is given by 
p-p Coulomb collisions and $\alpha$-p Coulomb collisions \citep[see Appendix A in][]{balbus08}.

In the X-ray binary disc model the plasma is cooler and less dense, 
and electrons are non-degenerate. These characteristics enhance the 
e-p collision probability, resulting in a resistivity that is 
several orders of magnitude higher than in neutrino cooled discs. 
The viscosity is also much higher, and moreover, it decreases 
towards larger radii. This is the opposite trend to that seen 
for the viscosity in neutrino cooled discs.

\section{The Magnetic Prandtl number}
\label{sec:pr}
%%%%%%%%%%%%%%%%%%%%%%%%%%%%%%%%%%%%%%%%%%%%%%%%%%%%%%%%%%%%%%%%%%%%%%%%%
\begin{figure}
\psfig{file=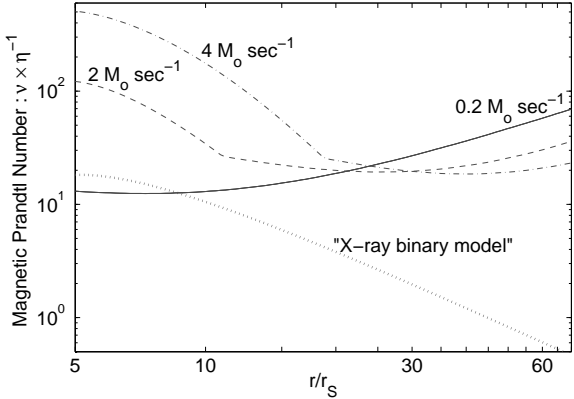,width=0.48\textwidth}
\caption[]{Magnetic Prandtl number as a function of scaled radius computed
for the three neutrino cooled disc models shown in
Figure~\ref{fig:T_r}.  
%The models are calculated for a black hole of
%mass $3 \ M_\odot$ and $\alpha = 0.1$. 
The magnetic Prandtl number for the "X-ray binary" 
disc model is also shown for comparison, as
a dotted curve.}
\label{fig:pr_r}
\end{figure}
%%%%%%%%%%%%%%%%%%%%%%%%%%%%%%%%%%%%%%%%%%%%%%%%%%%%%%%%%%%%%%%%%%%%%%%%%

With the results for the resistivity and viscosity in hand we can compute the magnetic Prandtl number, defined as 
the ratio of the microscopic viscosity to the resistivity,
\be
\Pr = \frac{\nu}{\eta}.
\label{eq:pr}
\ee Results for the three disc models that we have discussed in detail
are shown in Figure~\ref{fig:pr_r}. The radial behaviour of $\Pr$ is
mainly determined by the viscosity, since the resistivity is a rather
weak function of radius. In optically thin regions of the disc
(everywhere for $\dot{M}=0.2 \msunsec$, and at large radii, $r \gsim
15-30$ $\rs$, for the higher accretion rate models) the $\Pr$
number is relatively flat but eventually increases with radius because the rapidly decreasing density
increases the microscopic viscosity in the disc.  In the optically
thick regions, the temperature increases more rapidly toward the black
hole than does the density, and the increased viscosity leads to
larger $\Pr$ in the central parts.  We note that, although the values
of both the resistivity and the viscosity are far larger in the photon
cooled X-ray binary disc model, the value of the magnetic Prandtl
number is not all that different in these inner parts. More
significant is the different radial behaviour. For hyper-accreting
discs $\Pr$ increases with radius once the disc becomes optically
thin, and this trend continues throughout the neutrino cooled
zone. For X-ray binary discs $\Pr$ falls off with radius and
eventually becomes smaller than one (at $r\sim 45$ r$_{\rm g}$).
Our results for this model are qualitatively similar 
to those of \cite{balbus08} (see their fig.~1, left panel) with the quantitative differences 
arising due to different 
boundary conditions and value of $\eta_{\rm S}$. 

%%%%%%%%%%%%%%%%%%%%%%%%%%%%%%%%%%%%%%%%%%%%%%%%%%%%%%%%%%%%%%%%%%%%

\begin{figure}
\psfig{file=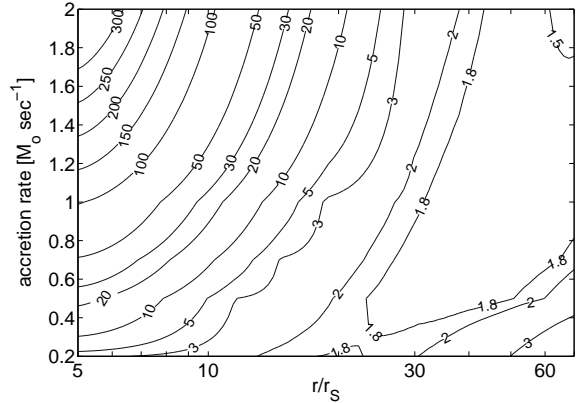,width=0.48\textwidth}
\caption[]{Contours of magnetic Prandtl number in
hyper-accreting discs as a function of radius and accretion rate,
computed for $\alpha=0.01$.}
\label{fig:pr_0.01}
\end{figure}
%-----------------------------------------------------------------------------
\begin{figure}
\psfig{file=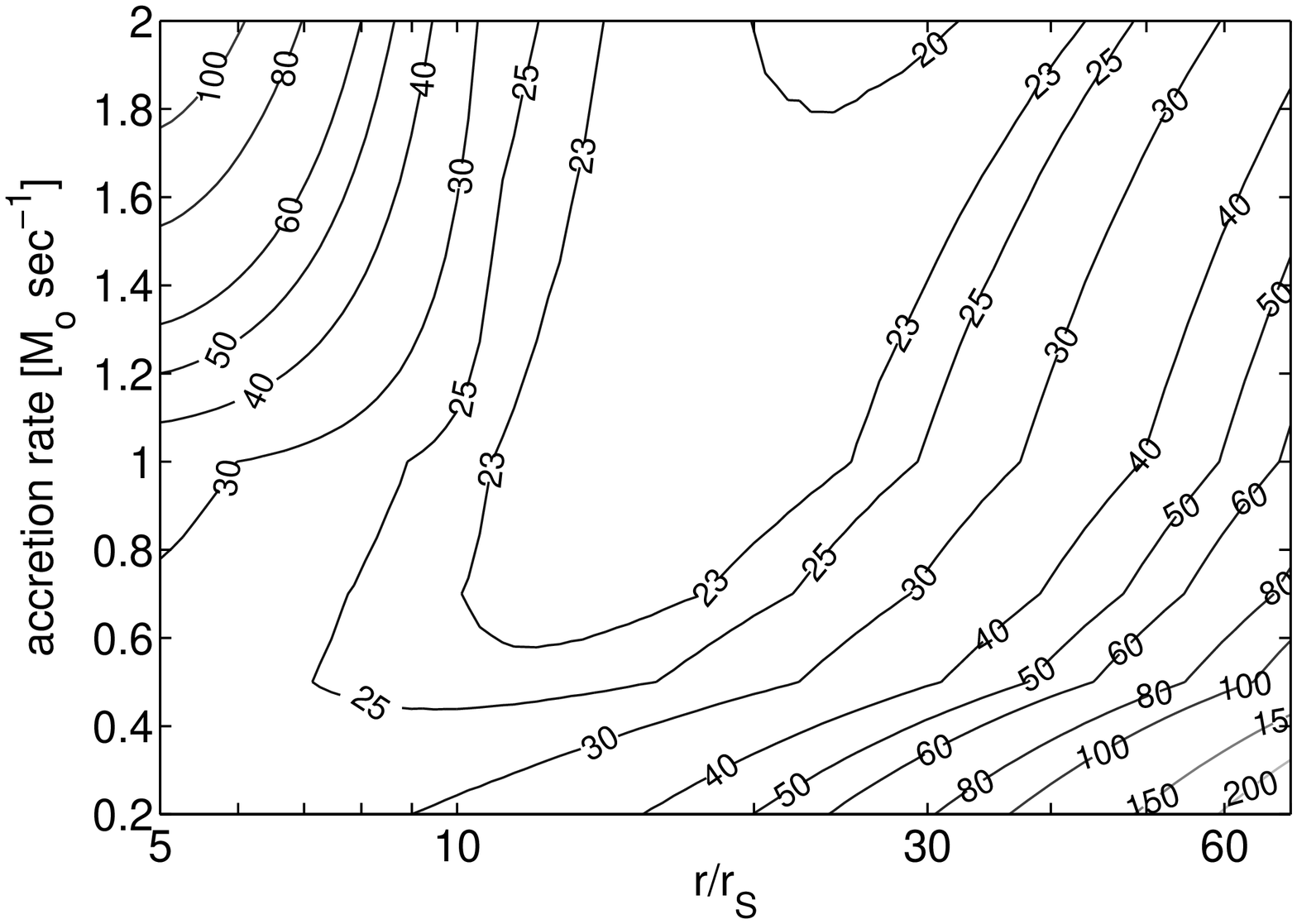,width=0.48\textwidth}
\caption[]{The same as Figure.~\ref{fig:pr_0.01}, but for $\alpha=0.1$.}
\label{fig:pr_0.1}
\end{figure}
%-----------------------------------------------------------------------------
\begin{figure}
\psfig{file=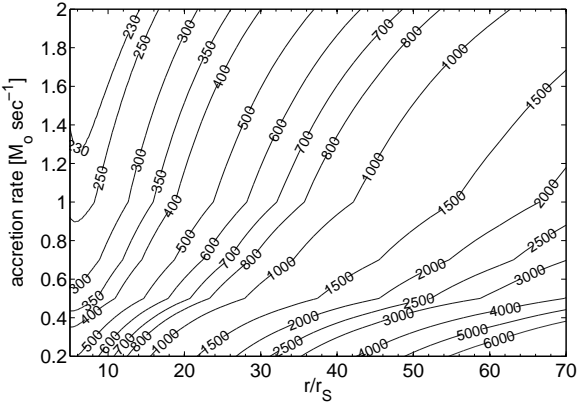,width=0.48\textwidth}
\caption[]{The same as Figure.~\ref{fig:pr_0.01}, but for $\alpha=1$.}
\label{fig:pr_1}
\end{figure}
%%%%%%%%%%%%%%%%%%%%%%%%%%%%%%%%%%%%%%%%%%%%%%%%%%%%%%%%%%%%%%%%%%%%

Figures~\ref{fig:pr_0.01},~\ref{fig:pr_0.1} and~\ref{fig:pr_1} show
the dependence of $\Pr$ on radius and accretion rate for three
different values of $\alpha$, $\alpha=0.01, 0.1~ {\rm and}~ 1$. 
We find that $\Pr >1$ throughout the
neutrino cooled region for all accretion rates in the range $0.2 \
M_\odot \ {\rm sec}^{-1} < \dot{M} < 2 \ M_\odot \ {\rm sec}^{-1}$.
Typically $\Pr \gg 1$ (except in the $r>20 \rs$ regions for $\alpha =0.01$), with
values as high as a few $10^3$ if $\alpha = 1$. 

As $\alpha$ decreases, the disc density is higher and the viscosity
drops: the lower viscosity results in a smaller magnetic Prandtl number.  
This is unlike X-ray binary discs, where the $\alpha$
dependence of the microscopic viscosity is largely offset by a similar
change in the resistivity \citep{balbus08}.

All the results discussed here are computed assuming midplane disc
conditions. Moving toward the upper layers of the disc the density
invariably drops more rapidly than the temperature, and as a
consequence the electrons become non-degenerate. However, verticle mixing is efficient and the
nuclear composition ($\yp$) at higher latitude is close to that at the midplane \citep{rossi07}. 
Under these conditions the resistivity increases more
slowly than the viscosity (whose main contribution will be eventually come from neutrons) 
and $\Pr$ increases further.

%------------------------------------------------------------------------------------------------------------
\section{Other non-ideal MHD effects}
Our motivation for computing the magnetic Prandtl number derives from
various suggestions that the saturation level of the MRI
varies strongly with $\Pr$. Almost all numerical simulations
addressing this issue include ohmic dissipation as the only non-ideal
term in the MHD equations. If other non-ideal terms are dominant in
hyper-accreting discs there will be (at the very least) additional
uncertainties in interpreting the results of simulations for
astrophysical discs\footnote{Most simulations to date have examined 
the $\Pr \gg 1$ regime in the absence of the Hall effect, while 
simulations that have included the Hall term have focused on the 
regime where the magnetic Reynolds number is low (as is appropriate 
for protoplanetary discs). Although the Hall term should not directly 
alter the dissipative properties of the fluid, we cannot rule out that 
the behavior of a high Prandl number disc with a strong Hall effect 
might differ from the pure $\Pr \gg 1$ case.}. Here, we estimate the magnitude
of the different non-ideal terms for fiducial values of temperature,
$T = 3\times 10^{10}$~K, and density, $\rho= 5\times 10^{10}$ gr
cm$^{-3}$. Under these conditions the plasma composition is $Y_{\rm p}
\simeq 5.5\times 10^{-2}$, the chemical potential is $\mu = 8.7$ and
the sound speed $c_{\rm s} = 1.6 \times 10^{9}$ cm s$^{-1}$. The
equipartition magnetic field, \be B_{\rm eq} =\sqrt{8 \pi \,P_{\rm b}}
\ee is $1.8 \times 10^{15}$ Gauss.

The low resistivity of plasma in neutrino cooled discs results in a 
very large magnetic Reynolds number. If we take $H =5.3 \times 10^{6}$~cm 
($\simeq 0.3 \times 20~\rs$), then

\be
R_{\rm eM}= \frac{c_{\rm s}H}{\eta} \simeq 1.5 \times 10^{19},
\label{eq:reynum}
\ee
where we have used the consistent value of the resistivity $\eta = 5.85 \times 10^{-4}$
cm$^{2}$ s$^{-1}$. {\em All} non-ideal MHD effects are therefore small. 
In relative terms, however, the high density means that the plasma 
is sufficiently collisional that, for very weak magnetic fields, 
ohmic dissipation is stronger than both the Hall effect (the separation of
charge carriers in the presence of a magnetic field perpendicular to
the current) and ambipolar diffusion (the differential motion 
of neutrons and charged particles). The dominance of ohmic dissipation 
persists until the magnetic
field becomes strong enough that the cyclotron frequencies of the charged
particles exceed the collision frequencies. We can write the relative 
strength of the ohmic (O) and Hall (Ha) effects as \citep[e.g.][]{sano02},
\be
\frac{\rm Ha}{\rm O} = \omega_{\rm ce} t_{\rm ep}. 
\label{eq:ho} 
\ee
where $\omega_{\rm ce}$ is the electron gyro-frequency and 
$t_{\rm ep}$ the e-p deflection timescale. For our conditions 
\be
t_{\rm ep} = \frac{\gamma_{e}}{4 \pi r_{\rm e} \eta n_{-}} \simeq
 \frac{\mu+1}{3.54 \times 10^{-12} \eta n_{p}} \simeq 2.87 \times 10^{-18} {\rm sec},
\label{eq:tep_2}
\ee
where the electron density has been evaluated as $n_{-} \simeq n_{p} = \rho {\yp}/ {\mp}$, 
and
\be
\omega_{\rm ce} = \frac{1.76 \times
10^{7}} {\left<\gamma_{\rm e}\right>} \frac{\rm B}{\rm gauss } \,{\rm Hz},
\ee 
where the mean Lorentz factor  is $\gamma_{\rm e}\simeq \mu+1$.
The field B here is the mean local field. Eq.~\ref{eq:ho} then gives, 
numerically, 
\be
 \frac{\rm Ha}{\rm O} = 5.2 \times 10^{-12} \frac{\rm B}{\rm gauss}.
\ee 
Likewise, the ratio of ambipolar (A) to Hall (H) strength can be 
written as \citep[e.g.][]{sano02},
\be
\frac{\rm A}{\rm Ha} = \omega_{\rm cp}\; t_{\rm pn}\;(2-2\yp) = 9.4 \times 10^{-17}  \frac{\rm B}{\rm gauss },
\label{eq:ah}
\ee
where $\omega_{\rm cp}=\omega_{\rm ce} (\gamma_e\me/\mp)$ is the proton gyro-frequency and the p-n collision time-scale is
$t_{\rm pn}=5.16\times 10^{-21}$ sec. We note again that the Coulomb coupling ($t_{\rm ep}$)
is weaker than the strong collision coupling for our plasma conditions.

Eq.~\ref{eq:ho} and eq.~\ref{eq:ah} show that ohmic dissipation is the
dominant non-ideal MHD term only in weakly magnetised plasmas which
have a ratio of thermal to magnetic energy $\beta > 10^8$. This means that,
for any reasonable magnetic field in the disc, 
the Hall effect is the dominant non-ideal MHD effect.
 For our
fiducial parameters the Hall effect becomes the most important
non-ideal MHD effect for $B\gsim 2 \times 10^{11} $
G\footnote{When $ \omega_{\rm ce} t_{\rm ep}\ge 1 $, $t_{\rm ep}$
and $\eta$ are formally the relaxation timescale and the resistivity
along the magnetic field lines. The transverse currents have a
collision timescale reduced by $1/(1+\omega_{\rm ce} t_{\rm ep})$.}.
At equipartition ambipolar diffusion is also more important than
resistive diffusion but we remain in the Hall regime.
%------------------------------------------------------------------------------------------------------------
\section{The radiatively inefficient zone}
\label{sec:pr_adaf}

In the outer disc --- beyond $\sim$ one hundred $\rs$ --- the physical
conditions are quite different. In this region the disc is
photon-pressure dominated. The temperature ($T\lsim 10^{10} {\rm K}$)
is below the neutrino cooling threshold and the flow is effectively
non-radiative. The free baryons are in the form of $H_{\rm e}$ nuclei
\citep{cb07} and the density ($\rho \lsim 10^{7} {\rm gr cm^{-3}}$) is
low enough that the electrons are non-degenerate (and non
relativistic). The disc is thus both more resistive and more viscous,
and the resistivity and viscosity have the classical Spitzer
dependencies on temperature and density (eq.~\ref{eq:etas} and
eq.~\ref{eq:nuspitzer}) but with numerical coefficients (also in the
``Spitzer'' Coulomb logarithms eqs.~\ref{eq:cou_log_ep_S}
and~\ref{eq:cou_log}) adjusted to be appropriate for a completely ionised He 
flow.  It is worth emphasizing that unlike the radiatively inefficient
flows discussed in the X-ray binary or AGN contexts
\citep[e.g][]{maha97,tanaka06}, those in hyper-accreting scenarios
would remain fully collisional. This constitutes a rare situation in which the 
collisionality guarantees a unique temperature for electrons and
nuclei. Moreover MHD dissipation scale arguments based on classical
Spitzer values are adequate.

The magnetic Prandtl number in this region is,

\be
\Pr = 7.3 \times 10^{-28} \frac{T^4}{\rho \Lambda},
\label{eq:pr_adaf}
\ee

\no
where $\Lambda = \log{\Lambda_{\rm eHe}} \times \log{\Lambda_{\rm HeHe}}$ is the product of the Coulomb logarithm for $e-H_{\rm e}$
and for $H_{\rm e}-H_{\rm e}$ collisions.

%\no
%Since $r\sim H$, 
%the radial density profile is
%\be
%\rho \simeq \frac{\dot{M}} {4\pi r^2 \alpha v_{\phi}} =\frac{1.9 \times 10^7}{r_{100}^{3/2}}\frac{\dot{M}_{0.2}}{M_{\rm bh,3}\;\alpha_{0.1}}~ {\rm gr~cm^{-3}},
%\label{eq:rhoadaf}
%\ee

If we assume that the temperature is set by a fraction of the
virial energy density then $a T^{4} = f~ (GM_{\rm bh}/r)\rho$. Numerically 
we find that $T^4/\rho = 5.9 \times 10^{34} f \left(\rs/r\right)$ and

\be
\Pr \simeq \frac{1.44 \times 10^6}{(\Lambda/15)} \left(\frac{\rs}{r}\right),
\label{eq:pr_adaf}
\ee

%\be
%T = f^{1/4} \frac{10^{10}}{r_{100}^{5/8}} \left(\frac{\dot{M}_{0.2}}{\alpha_{0.1}}\right)^{1/4} M_{\rm bh,3}^{-1/2}~ {\rm K},
%\label{eq:Tadaf}
%\ee

\no where we take $f=1/2$~\footnote{We are aware that there is no
consensus on the exact value of $f$; however, the dependence of $\Pr$
on $f$ is small.}. %In eqs.~\ref{eq:rhoadaf} and eqs.~\ref{eq:Tadaf},
%the accretion rate is normalised to $0.2 \msunsec$, the BH mass to $3 M_{\sun}$,
%$\alpha$ to $0.1$ and the radius to $100 r_{\rm s} = 8.9 \times 10^7 M_{\rm bh,3}~ {\rm cm}$.
$\Lambda$ varies very slowly with radius and eq.~\ref{eq:pr_adaf} with
  $(\Lambda/15)=1$ is actually a very good approximation for $\Pr$ in
  the whole outer disc.  We note that $\Pr \gg 1$ and it falls off
  slowly as $r^{-1}$: for $r \gsim 100 \rs$, $\Pr \lsim 10^4$.

This result indicates that in the outer regions of hyper-accreting
discs $\Pr$ has a very different behaviour than in the innermost
parts.  Here $\Pr$ is rather insensitive to the actual accreting
conditions $\dot{M}$ and $\alpha$ (as long as the outer disc remains
photon-pressure dominated), but it is a {\it stronger} function of
temperature and it decreases with radius.  However, both regions have
in common $\Pr >1$: unlike X-ray binary and AGN discs, hyper-accreting
discs seem not to have a $\Pr <1$ region.  Beside a high $\Pr$, this
outer disc flow has other characteristics in common with the neutrino-cooled
region.  The conductivity is high with $R_{\rm eM} \sim
10^{12}-10^{11}$ and the Hall effect dominates even for modest magnetic
fields ($\beta < 10^6-10^5$). These features are caused by the still
high temperature that makes Coulomb coupling rather inefficient (though in the 
outer region it is still more
efficient than nuclear coupling).

%------------------------------------------------------------------------------------------------------------
\section{Discussion}
\label{sec:discussion}  

In this section, we attempt to relate the plasma properties that we have 
computed to the global properties of discs.
There are many steps required to connect the
microscopic scales where resistivity and viscosity act to the global
evolution of the disc, and simulations and analytic arguments 
provide only an imperfect guide. However, plausible guesses are possible.
We focus on the possible consequences of  magnetic
Prandtl number greater than unity 
for the magnetisation and evolution of hyper-accreting discs.

Let us first summarise the argument that suggests that the magnetic
Prandtl number matters.  In turbulent plasmas with $\Pr > 1$, the
absence of turbulence below the viscous scale $\lambda_{\nu} \sim
R_{\rm e}^{-3/4} H$ means that the resistive scale may not be
easily accessible to field lines and reconnection could be
slow with respect to magnetic amplification. We can understand
the dependence on $\Pr$ with the following simple argument.
Turbulence at scale $\lambda_{\nu}$ exponentially amplifies the
magnetic field on an eddy turnover time-scale $t_{\rm eddy} = H/c_{\rm
s} \left(\lambda_{\nu}/H \right)^{2/3}$, while the magnetic field
dissipation on that same scale takes $t_{\eta} =
\lambda_{\nu}^{2}/\eta$. The ratio of these timescales is \be
\frac{t_{\eta}}{t_{\rm eddy}} \simeq \Pr.  \ee So, for $\Pr > 1$,
magnetic energy can build up, first on the viscous scale and
subsequently in an upward cascade toward larger scales
\citep{brandenburg01}.  

This may have important consequences for
accretion discs. Within discs, the turbulent growth of magnetic fields
can be limited -- in principle -- by either resistive dissipation (on
small scales) or buoyant expulsion of the flux from the
disc. Numerical simulations suggest that in low $\Pr$
discs resistive dissipation ultimately sets the saturation level
\citep[e.g.][]{stone96}. If, however, the small-scale dissipation of
magnetic fields is frustrated by a high magnetic Prandtl number, the
only channel available to limit the flux may be macroscopic
buoyancy. Since buoyancy is relatively inefficient -- the timescale is
likely to modestly exceed the Alfv\'en crossing time $H/V_{\rm A}$ of
the disc -- a high magnetic Prandtl number disc could
saturate at a level that involves stronger magnetic fields, more
vigorous turbulence and angular momentum transport, and ongoing field
expulsion. These expectations are supported by the available numerical
evidence. \cite{lesur07} and \cite{fromang07} find that the parameter
$\alpha$ increases with the magnetic Prandtl number in the range $2
\le \Pr \le 8$ and $R_{\rm e}\sim 10^{3}-10^{4}$. In the case where
$\alpha$ approaches unity, \cite{lesur07} find a highly time dependent
disc, with large fluctuations in $\alpha$ and the mean magnetic
pressure.  This appears to be a consequence of the well-known
dependence of the strength of the MRI on the magnetic field: as the
ratio $\beta$ of thermal to magnetic pressure decreases $\alpha$ at
first increases until the MRI is quenched around equipartition
\citep{hawley95,lesur07}.

Based on these arguments, two evolutionary consequences can be
contemplated for hyper-accreting discs formed after stellar core
collapse or compact object mergers. In the initial phase, a high
magnetic Prandtl number in the neutrino cooled inner disc results in
stronger magnetic field and a higher efficiency of angular momentum
transport. Since, as we have shown, a {\it higher} $\alpha$ in such
discs results in a yet {\it higher} $\Pr$, 
a runaway occurs until $\alpha \sim 1$. This regime is sustainable
for many dynamical timescales, provided that the excess magnetic
energy that could quench the MRI is removed from the disc, for
example, by buoyancy. This is plausible since close to equipartition
the buoyancy timescale $\sim H/V_{\rm A} \sim \Omega^{-1}$ is similar
to the orbital timescale and can provide cooling by removing magnetic
energy\footnote{Note that we investigated a disc structure with
$\alpha=1$ in which a fixed fraction of the cooling occurs
non-radiatively via loss of magnetic energy. We found that $\Pr$ is
only reduced by a factor of $\sim 2$ with respect to a disc with
$\alpha=1$ in which only neutrino cooling and gas pressure are
considered. Hence reasonable amounts of non-radiative cooling
do not avert a runaway to a high $\alpha$ state.}. The disc,
which would be in a similar regime to that simulated by
\cite{lesur07}, would be highly variable. This initial stage, in which
a highly variable magnetic ``corona" is built above the disc via flux
expulsion, could be associated with the production of a magnetic
outflow, dynamically similar to those studied by \citet{proga03} and
\citet{mckinney06}  \citep[see also][]{king04}, polluted with some disk matter, that eventually will
be observed as a GRB.

The longer term evolution of very high $\Pr$ discs is less clear. One 
possibility is that eventually a {\em net} vertical field, which 
cannot be removed locally by buoyancy, builds up and suppresses the MRI. 
This depends upon the
ability of field lines to diffuse outwards. 
Turbulent diffusion can depend in turn on the strength of the magnetic field
and on non-linear MHD effects. In our case the dominance of the Hall effect may 
influence the behaviour of the turbulence, by introducing an asymmetry between magnetic fields
that are aligned or anti-aligned with the direction of the angular momentum vector \citep{sano02}.
The outcome is thus uncertain and the investigation requires numerical methods.

 From a purely phenomenological perspective, one might note that there are similarities between
the state transitions in X-ray binaries and GRB prompt and late-time activity.
Our results show the presence of a high $\Pr$ number inner region in both discs.
One may then speculate that the GRB prompt phase could involve similar physical 
conditions, insofar the magnetic field behaviour is concerned, to the transition to the ``low/hard state'' in X-ray 
binaries, which is observationally associated with the emergence of an outflow \citep{fender99}.
However, the theoretical analogy with X-ray binaries is imperfect,
since it is unclear that what causes a change in state may be the same physical mechanism
in both sources.
For X-ray binary discs the temperature dependence of $\Pr$ may allow a local
thermal runaway, and it has been suggested that this ultimately causes
global state transitions \citep{balbus08}. The conditions 
for such a mechanism to work in hyper-accreting discs are present only in the outer 
radiatively inefficient regions,
since in the neutrino-cooled regions $\Pr$ is primarily a
function of density. However, a classical thermal instability 
is not possible in radiatively inefficient flows.

\section{Conclusions}

In this paper, we calculated the magnetic Prandtl number for plasma
conditions in hyper-accreting discs. This dimensionless
number $\Pr = \nu/\eta = \lambda_{\nu}^{2}/\lambda_{\eta}^{2}$
expresses the relative size of two critical turbulent flow
scales: the scale on which velocity fluctuations are viscously damped
$\lambda_{\nu}$, and the scale $\lambda_{\eta}$ on which ohmic losses
result in magnetic field dissipation. We find that in the inner 
neutrino-cooled regime:
\begin{itemize}
\item[(1)]
Electric resistivity involves
relativistic, mildly degenerate electrons. As a consequence,
the resistivity is very low and weakly dependent on density and temperature.
\item[(2)]
The main source of viscosity is Coulomb collisions between mildly {\it degenerate electrons} and  protons.
It decreases as the flow gets denser or cooler, since the electron fraction decreases.
\item[(3)]
The magnetic Prandtl number is always {\em greater} than unity, unless the angular momentum transport 
efficiency, parameterized via the Shakura-Sunyaev $\alpha$, is 
very small ($\alpha < 0.01$). $\Pr$ ranges between a few tens 
to a few $10^3$, and it typically increases with $\alpha$.
In the optically {\em thin} regions for neutrinos, the main dependence of $\Pr$ is on density 
(inversely, via viscosity)
since the disc temperature varies very slowly as a function of $\dot{M}$ and radius.
This causes $\Pr$ to {\em decrease} with increasing accretion rates.
In the optically {\em thick} regions, the main dependence is on temperature, that varies
more rapidly than density. As a consequence, $\Pr$ {\em increases} with $\dot{M}$
\end{itemize}

\no In the outer, radiatively inefficient region:

\begin{itemize}
\item[(4)] Resistivity and viscosity have the classical ``Spitzer'' dependencies on temperature and density,
since electrons are non-relativistic and non-degenerate and viscosity is mainly given by Coulomb interactions 
between ionised helium particles.
The magnetic Prandtl number is thus a much stronger function of temperature $\Pr \propto T^{4}/\rho$.
In contrast, it does not depend on the overall magnitude of accretion (via $\dot{M}$ and $\alpha$). The 
Prandtl number is always much larger
than unity ($\Pr \simeq 10^4 (100~\rs)/r$) and it is unlikely to fall below unity for the entire
 extent of the disc.
\end{itemize}

\no We conclude by noting a number of implications for the global structure of 
hyper-accreting discs:

\begin{itemize}
\item[(5)] For all plausible values of the magnetic field strength in the 
disc, the Hall effect provides the largest non-ideal term 
in the MHD equations. 
These discs are perhaps a unique example of a high magnetic Reynolds 
number flow in which the 
Hall effect is the dominant non-ideal term. Since there are no simulations 
where these conditions are accounted for,
their effect on the magnetic field evolution and MRI is unclear.
\item[(6)] The large values of the magnetic Prandtl number mean that
the evolution of magnetic field within hyper-accreting discs is likely
to be qualitatively different from discs with
lower magnetic Prandtl numbers. The results of forced dynamo and MRI
simulations suggest that, in the high $\Pr$ regime, small scale field
dissipation is suppressed and the saturation level of the magnetic
field is enhanced, perhaps dramatically
\citep{lesur07,fromang07}. Numerical simulations of hyper-accreting
discs that do not account for microphysical dissipation may therefore
have underestimated the magnetic field strength.
\item[(7)]
The {\em immediate} consequence of a high value of $\Pr$ for 
models of the central engines of GRBs is that 
disc formation is likely to be accompanied 
by vigorous expulsion of magnetic flux. This favours models 
in which energy is liberated in the form of a magnetic jet 
or outflow.
\end{itemize}

%------------------------------------------------------------------------------------------------------------
%------------------------------------------------------------------------------------------------------------

\section*{Acknowledgments}
The authors thank A. Potekhin for providing help with his code and for useful discussions.
They also acknowledge useful discussion with C. Thompson, J. McKinney, Y. Levin and A. Beloborodov.  
EMR acknowledges support from NASA though Chandra Postdoctoral Fellowship grant number 
PF5-60040 awarded by the Chandra X-ray Center, which is operated by the Smithsonian 
Astrophysical Observatory for NASA under contract NASA8-03060. PJA acknowledges support from NASA 
under grants NNG04GL01G and NNX07AH08G from the Astrophysics Theory Program.
%------------------------------------------------------------------------------------------------------------
%------------------------------------------------------------------------------------------------------------


\begin{thebibliography}{}

\bibitem[Arav \& Begelman(1992)]{arav92}
Arav, N. \& Begelman, M.C. 1992, \apj 401, 125

\bibitem[Balbus \& Hawley(1998)]{balbus98}
 Balbus, S. A., Hawley, J. F. 1998, Reviews of Modern Physics, 70, 1

\bibitem[Balbus \& Henri(2008)]{balbus08}
 Balbus, S. A., \& Henri, P. 2008, \apj, 674, 408

\bibitem[Bethe(1949)]{bethe}
Bethe, H.A. 1949, Phys.Rev., 76, 38

%\bibitem[Ball, Narayan \& Quataert(2001)]{ball01}
%Ball, G.~H., Narayan, R., \& Quataert, E. 2001, \apj, 552, 221

\bibitem[Beloborodov(2003)]{belo03}
Beloborodov, A.~M. 2003, \apj,588, 944


\bibitem[Blandford \& Payne(1982)]{blandford82}
 Blandford, R.~D., Payne, D.~G. 1982, MNRAS, 199, 883

\bibitem[Brandenburg(2001)]{brandenburg01}
 Brandenburg, A. 2001, ApJ, 550, 824

\bibitem[Braginskii(1957)]{bra57}
Braginskii, S.~I. 1957, J.Exptl.Theoret.Phys.,33,459

\bibitem[Braginskii(1958)]{bra58}
Braginskii, S.~I. 1958, Soviet Phys. JETP, 6,358

\bibitem[Burrows et al.(2005)]{burrows05}
Burrows, D.~N. et al. 2005, Sci, 309, 1833


\bibitem[Chen \& Beloborodov(2007)]{cb07}
Chen, W-X, \& Beloborodov A.M. 2007, \apj, 657:383

\bibitem[Di Matteo, Perna \& Narayan(2002)]{dimatteo02}
Di Matteo, T., Perna, R., Narayan, R., 2002, \apj, 579, 706

\bibitem[Fender et al.(1999)]{fender99}
Fender, R. et al. 1999,\apj, 519, 165 

\bibitem[Fromang et al.(2007)]{fromang07} 
 Fromang, S., Papaloizou, J., Lesur, G., \& Heinemann, T. 2007, 
 A\&A, 476, 1123

%\bibitem[Janiuk et al.(2004)]{janiuk04}
%Janiuk, A., Perna, R., Yuan, Y., Di Matteo, T. \& Czerny, B. 2007, \mnras, 355, 950 

\bibitem[Hackenburg(2006)]{hack06}
Hackenburg, H., 2006, PRC, 73, 044002


\bibitem[Hawley, Gammie \& Balbus(1995)]{hawley95}
Hawley, J. F., Gammie, C. F., \& Balbus, S. A. 1995, \apj, 440, 742

%\bibitem[Janiuk et al.(2007)]{janiuk07}
%Janiuk, A., Yuan, Y., Perna, R. \& Di Matteo, T. 2007 \apj, 664, 1011 

\bibitem[King et al.(2004)]{king04}
King,  A. R., Pringle,  J. E.,  West, R. G.,  Livio, M., 2004, \mnras, 348, 111


%\bibitem[Koers \& Giannios(2007)]{koers07}
%Koers, H.~B.~J., \& Giannios, D. 2007 \aap, 471, 395

\bibitem[Kohri \& Mineshige(2002)]{kohri02}
Kohri, K., \& Mineshige, S. 2002, \apj, 577, 311


%\bibitem[Lee, Ramirez-Ruiz \& Page(2004)]{lee04}
%Lee, W.H., Ramirez-Ruiz, E. \& Page, D. 2004, \apj, 608, 5

\bibitem[Lesur \& Longaretti(2007)]{lesur07}
 Lesur, G., \& Longaretti, P.-Y. 2007, MNRAS, 378, 1471

\bibitem[Lynden-Bell \&  Pringle(1974)]{lynden74}
Lynden-Bell, D. \& Pringle, J.~E. 1974, \mnras, 168, 603

\bibitem[Longair(1992)]{longair}
 Longair, M.~S., ``High Energy Astrophysics'', 1992, Cambridge University Press


\bibitem[Mahadevan \& Quataert(1997)]{maha97}
Mahadevan, R. \& Quataert, E., 1997, \apj, 490, 605


\bibitem[Masada et al.(2007)]{masada07}
Masada, Y., Kawanaka, N., Sano, T., \&  Shibata, K. 2007, \apj, 663, 437

\bibitem[McKinney(2006)]{mckinney06}
McKinney, J. C. 2006, \mnras, 368, 1561

%\bibitem[Meyer \& Meyer-Hofmeister(1981)]{meyer81} 
%Meyer, F., Meyer-Hofmeister, E. 1981, \aap, 104, L10

\bibitem[Miller et al.(1990)]{miller90} 
Miller,G. A.,Nefkens, B.M.K., \& \v{S}laus, I. 1990, PhR, 194, 1

\bibitem[Nandkumar \& Pethick(1984)]{nand84}
Nandkumar, R. \& Pethick, C.~J., \mnras, 209, 511

\bibitem[Narayan, Piran \& Kumar(2001)]{narayan01}
Narayan, R., Piran, Tsvi, \& Kumar, P., 2001, \apj, 557, 949

\bibitem[Popham, Woosley \& Fryer(1999)]{popham99}
 Popham, R., Woosley, S. E., \& Fryer, C. 1999, \apj, 581,356

\bibitem[Potekhin(1999)]{pot99a}
Potekhin, A.Y 1999, \aap, 351, 797

\bibitem[Potekhin et al.(1999)]{pot99b}
Potekhin, A.Y, Baiko, D.A., Haensel, P., \& Yakovlev, D.G. 1999, \aap, 351, 797

\bibitem[Pringle(1981)]{pringle81}
 Pringle, J. E. 1981, ARA\&A, 19, 137

\bibitem[Proga et al.(2003)]{proga03}
Proga, D., MacFadyen, A. I., Armitage, P. J., \& Begelman, M. C. 2003, \apj, 599, L5

\bibitem[Rossi, Armitage \& Di Matteo(2007)]{rossi07}
Rossi, E.~M., Armitage, P.~J. \& Di Matteo, T., 2007, \apss, 311, 185


\bibitem[Ruffert \& Janka(1999)]{ruffert99}
 Ruffert, M. \& Janka, H.-T.1999, \aap, 344, 573

\bibitem[Sano \& Stone (2002)]{sano02}
Sano, T., \& Stone, J.~M., 2002, \apj, 570, 314

\bibitem[Schekochihin et al.(2004)]{schekochihin04}
 Schekochihin, A. A., Cowley, S. C., Taylor, S. F., Maron, J. L., \& McWilliams, J. C. 2004, 
 ApJ, 612, 276

\bibitem[Shakura \& Sunyaev(1973)]{shakura73}
 Shakura, N. I., \&  Sunyaev, R. A. 1973, A\&A, 24, 337

\bibitem[Spitzer(1962)]{spitzer62}
 Spitzer, L. 1962, Physics of Fully Ionized Gases, New York: Interscience (2nd edition)

\bibitem[Spruit, Stehle \& Papaloizou 1995]{spruit95}
Spruit, H.~C., Stehle, R., \& Papaloizou, J.~C.~B. 1995, \mnras, 275, 1223

%\bibitem[Stepney(1983)]{stepney83}
%Stepney,S.1983, \mnras, 202, 467

\bibitem[Stone et al. (1996)]{stone96}
Stone, J.M., Hawley, J.F., Gammie, C.F., \& Balbus, S.A. 1996, \apj, 463,656

\bibitem[Tanaka \& Menou(2006)]{tanaka06}
Tanaka, T., \& Menou, K., 2006, \apj, 649, 345

\bibitem[Umurhan, Menou \& Regev(2007a)]{umurhan07} 
 Umurhan, O. M., Menou, K., \& Regev, O. 2007a, PRL 98, 034501

\bibitem[Umurhan, Regev \& Menou(2007b)]{umurhan07b} 
 Umurhan, O. M., Regev, O. \& Menou, K. 2007b, PRE 76, 036310


\bibitem[Woosley(1993)]{woosley93}
 Woosley, S. E. 1993, ApJ, 405, 273

%\bibitem[Yao et al.(2006)]{yao06}
%Yao, W.-M. et al. 2006, JPhG, 33, 1



\end{thebibliography}
\end{document}